\documentclass[12pt]{article} 

\usepackage{latexsym} 	
\usepackage{amssymb}  	
\usepackage{amsbsy}
\usepackage{epsfig}     
\usepackage{graphics,color}
\usepackage{a4}

\newcommand{\eps}{\epsilon}

\newcommand{\Tr}{\mbox{Tr}\,}
\newcommand{\id}{1\!\!1}

\newcommand{\cP}{{\cal P}}
\newcommand{\bra}{\langle}
\newcommand{\ket}{\rangle}

\newcommand{\half}{\frac{1}{2}}
\newcommand{\re}{\mbox{Re\,}}
\newcommand{\im}{\mbox{Im\,}}

\newcommand{\xv}{{\mathbf x}}

\newcommand{\be}{\begin{equation}}
\newcommand{\ee}{\end{equation}}
\newcommand{\bea}{\begin{eqnarray}}
\newcommand{\eea}{\end{eqnarray}}
\newcommand{\bean}{\begin{eqnarray*}}
\newcommand{\eean}{\end{eqnarray*}}
\newcommand{\nn}{\nonumber}
\newcommand{\hm}{\hspace*{-0.6cm}}
\newcommand{\diag}{\mbox{diag}}

\newcommand{\bit}{\begin{itemize}}
\newcommand{\eit}{\end{itemize}}

\renewcommand{\theequation}{\arabic{section}.\arabic{equation}}

\begin{document}

\title{ 
\vskip -100pt
{\begin{normalsize}
\mbox{} \hfill arXiv:0807.1597 [hep-lat] \\
\vskip  100pt
\end{normalsize}}
\bf\Large
  Stochastic quantization at finite chemical potential 
}
 
\author{
 \addtocounter{footnote}{2}
 Gert Aarts\thanks{email: g.aarts@swan.ac.uk}
 $\;$
 and
 Ion-Olimpiu Stamatescu\thanks{email: 
 I.O.Stamatescu@thphys.uni-heidelberg.de} 
 \\ \mbox{} \\
 $^\ddag${\em\normalsize Department of Physics, Swansea University,
 Swansea, United Kingdom}
 \\
 $^\S${\em\normalsize Institut f\"ur Theoretische Physik, 
 Universit\"at Heidelberg, Heidelberg, Germany} \\
 {\em\normalsize and FEST, Heidelberg, Germany}
}

 \date{July 10, 2008}

 \maketitle

 \begin{abstract}

 A nonperturbative lattice study of QCD at finite chemical potential is 
complicated due to the complex fermion determinant and the sign problem. 
Here we apply the method of stochastic quantization and complex Langevin 
dynamics to this problem. We present results for U(1) and SU(3) one link 
models and QCD at finite chemical potential using the hopping expansion. 
The phase of the determinant is studied in detail. Even in the region 
where the sign problem is severe, we find excellent agreement between the 
Langevin results and exact expressions, if available. We give a partial 
understanding of this in terms of classical flow diagrams and eigenvalues 
of the Fokker-Planck equation.

 \end{abstract}

\newpage


\tableofcontents


\section{Introduction}
\label{sec:Introduction}
\setcounter{equation}{0}

A nonperturbative understanding of QCD at nonzero baryon density remains 
one of the outstanding problems in the theory of strong interactions. 
Besides the theoretical challenge, there is a clear phenomenological 
interest in pursuing these studies, due to the ongoing developments in 
heavy ion collision experiments, at RHIC, LHC and the planned FAIR 
facility at GSI. 

The standard nonperturbative tool to study quarks and gluons, lattice QCD, 
cannot be applied in a straightforward manner, because the complexity of 
the fermion determinant prohibits the use of approaches based on importance 
sampling. This is commonly referred to as the sign problem. Since the start 
of the millenium a number of new methods has been devised. These include 
reweighting \cite{Fodor:2001au,Fodor:2001pe,Fodor:2002km,Fodor:2004nz}, 
Taylor series expansion in $\mu/T$ 
\cite{Allton:2002zi,Allton:2003vx,Allton:2005gk,Gavai:2003mf}, imaginary 
chemical potential and analytical continuation \cite{de Forcrand:2002ci,de 
Forcrand:2003hx,deForcrand:2006pv,D'Elia:2002gd}, and the use of the 
canonical ensemble \cite{Kratochvila:2005mk,Ejiri:2008xt} and the density 
of states \cite{Fodor:2007vv}. Except for the last two, these approaches 
are approximate by construction and aimed at describing the QCD phase 
diagram in the region of small chemical potential and temperatures around 
the crossover between the confined and the deconfined phase.
 In this paper we discuss an approach which is manifestly independent of 
those listed above: stochastic quantization and complex Langevin dynamics. 
How well this method will work is not known a priori. However, one of the 
findings of our study is that excellent agreement is found in the case of 
simple models, where comparison with results obtained differently is 
available. In particular we find that the range of applicability is not 
restricted to small chemical potential and, importantly, does not depend 
on the severity of the sign problem. The first results we present for 
lattice QCD at nonzero density are encouraging, although a systematic 
analysis has not yet been performed.

This paper is organized as follows. In the following section we briefly 
describe the idea behind stochastic quantization and the necessity to use 
complex Langevin dynamics in the case of nonzero chemical potential. In 
Secs.\ \ref{sec:onelink} and \ref{sec:su3onelink} we apply this technique 
to U(1) and SU(3) one link models. In both cases a comparison with exact 
results can be made. We study the phase of the determinant in detail. In 
the case of the U(1) model, we employ the possibility to analyse classical 
flow diagrams and the Fokker-Planck equation to gain further understanding 
of the results. In Sec.\ \ref{sec:polyakov} we turn to QCD, using the full 
gauge dynamics but treating the fermion determinant in the hopping 
expansion. Our findings and outlook to the future are summarized in Sec.\ 
\ref{sec:outlook}. The Appendix contains a brief discussion of 
the Fokker-Planck equation in Minkowski time for the one link U(1) model.

\section{Stochastic quantization and complex Langevin dynamics}
\label{sec:stoch}
\setcounter{equation}{0}

The main idea of stochastic quantization 
\cite{Parisi:1980ys,Damgaard:1987rr} is that expectation values are 
obtained as equilibrium values of a stochastic process. To implement this, 
the system evolves in a fictitious time direction $\theta$, subject to 
stochastic noise, i.e.\  it evolves according to Langevin dynamics.
 Consider for the moment a real scalar field $\phi(x)$ in $d$ 
dimensions with a real euclidean action $S$. The Langevin equation reads
\be
\label{eq:langphi}
 \frac{\partial\phi(x,\theta)}{\partial\theta} = -\frac{\delta 
S[\phi]}{\delta 
\phi(x,\theta)} + \eta(x,\theta),
\ee
where the noise satisfies
\be
 \bra \eta(x,\theta)\ket = 0, \;\;\;\;\;\;\;\;
 \bra \eta(x,\theta)\eta(x',\theta')\ket = 2\delta(x-x')
\delta(\theta-\theta').
\ee
By equating expectation values, defined as
\be
\bra O[\phi(x,\theta)]\ket_\eta = \int D\phi \, P[\phi,\theta]O[\phi(x)],
\ee
 where $O$ is an arbitrary operator and the brackets on the left-hand side 
denote a noise average, it can be shown that the probability distribution 
$P[\phi,\theta]$ satisfies the Fokker-Planck equation
 \be
\frac{\partial P(\phi,\theta)}{\partial \theta} = 
\int d^dx\, \frac{\delta}{\delta\phi(x,\theta)}\left( 
\frac{\delta}{\delta \phi(x,\theta)} + \frac{\delta S[\phi]}{\delta 
\phi(x,\theta)}\right) P[\phi,\theta].
\ee
 In the case of a real action $S$, the stationary solution of the 
Fokker-Planck equation, $P[\phi] \sim \exp\left(-S[\phi]\right)$, will be 
reached in the large time limit $\theta\to \infty$, ensuring convergence 
of the Langevin dynamics to the correct equilibrium distribution.
 When the action is complex, as is the case in QCD at finite chemical 
potential, the situation is not so easy. It is still possible to consider 
Langevin dynamics based on Eq.\ (\ref{eq:langphi})
\cite{Parisi:1984cs,Klauder:1985a,Klauder:1985b,Gausterer:1986gk}.
However, due to the 
complex force on the right-hand side, fields will now be complex as well: 
$\phi\to \re\phi+i\im\phi$. As a result, proofs of the convergence towards 
the (now complex) weight $e^{-S}$ are problematic.

In the past, complex Langevin dynamics has been applied to effective 
three-dimensional spin models with complex actions, related to lattice QCD 
at finite $\mu$ in the limit of strong coupling and large fermion mass 
\cite{Karsch:1985cb,Ilgenfritz:1986cd,Bilic:1987fn} (for applications to 
other models, see e.g.\ Ref.\ \cite{Ambjorn:1986fz}).
Our work has also partly been inspired by the recent application of stochastic 
quantization to solve nonequilibrium quantum field dynamics 
\cite{Berges:2005yt,Berges:2006xc,Berges:2007nr}. 
In that case the situation is even more severe. Nevertheless, numerical 
convergence towards a solution can be obtained under certain conditions,
both for simple models and four-dimensional field theories. For an  
illustration we present some original results in the appendix.

Here we consider models with a partition function whose form is motivated 
by or derived from QCD at finite chemical potential. The QCD partition 
function reads
\be
Z = \int DU\, e^{-S_B} \det M,
\ee
where $S_B(U)$ is the bosonic action depending on the gauge links $U$ and 
$\det M$ is the complex fermion determinant, satisfying
\be 
\label{eqdetsym}
\det M(\mu) = [\det M(-\mu)]^*. 
\ee
Specifically, for Wilson fermions the fermion matrix has the 
schematic form
\be
\label{eqMQCD}
M = 1 - \kappa\sum_{i=1}^3 \left( \Gamma_{+i} U_{x,i}T_{i} 
+ \Gamma_{-i} U_{x,i}^{\dagger} T_{-i} \right)
-\kappa \left( e^\mu \Gamma_{+4} U_{x,4}T_{4} +
e^{-\mu}\Gamma_{-4} U_{x,4}^{\dagger} T_{-4} \right).
\ee
 Here $T$ are lattice translations, $\Gamma_{\pm \mu} = \id\pm 
\gamma_\mu$, and $\kappa$ is the hopping parameter. Chemical potential is 
introduced by multiplying the temporal links in the forward (backward) 
direction with $e^{\mu}$ ($e^{-\mu}$) \cite{Hasenfratz:1983ba}. We use 
Eq.\ (\ref{eqMQCD}) as a guidance to construct the U(1) and SU(3) one link 
models considered next.

\section{One link U(1) model}
\label{sec:onelink}
\setcounter{equation}{0}

\subsection{Complex Langevin dynamics}

We consider a one link model with one degree of freedom, written as 
$U=e^{ix}$. The partition function is written suggestively as
 \be
 Z = \int dU \, e^{-S_B}  \det M
 = \int_{-\pi}^{\pi}\frac{dx}{2\pi}\, e^{-S_B}  \det M,
\ee
where the ``bosonic'' part of the action reads
 \be
 S_B(x) = -\frac{\beta}{2}\left( U+U^{-1} \right) = -\beta\cos x,
 \ee
 while the ``determinant'' is constructed by multiplying the forward 
(backward) link with $e^{\mu}$ ($e^{-\mu}$), 
 \be
 \det M = 1 + \half\kappa \left[ e^{\mu}U + e^{-\mu} U^{-1} \right] 
 = 1+\kappa \cos(x-i\mu).
\ee
 Due to the chemical potential, the determinant is complex and 
has the same property as the fermion determinant in QCD, i.e.\ 
$\det M(\mu) = [\det M(-\mu)]^*$. For an imaginary chemical potential 
$\mu=i\mu_I$, the determinant is real, as is the case in QCD.

Observables are defined as
\be
 \bra O(x)\ket = \frac{1}{Z} 
 \int_{-\pi}^{\pi}\frac{dx}{2\pi}\, e^{-S_B}
 \det M \, O(x).
\ee
 In this model most expectation values can be evaluated analytically. We 
consider here the following observables:
\bit
 \item
 Polyakov loop:
\be
 \bra U \ket = \bra e^{ix} \ket = \frac{1}{Z}\left[ I_1(\beta) + 
\kappa I_1'(\beta)\cosh\mu 
 -\kappa I_1(\beta)/\beta \sinh\mu \right],
\ee
 where the partition function equals
 \be 
 Z = I_0(\beta) + \kappa I_1(\beta) \cosh\mu,
 \ee 
and $I_n(\beta)$ are the modified Bessel functions of the first kind.
 \item
 Conjugate Polyakov loop:
\be
 \bra U^{-1} \ket = \bra e^{-ix} \ket = \bra e^{ix} \ket \bigg|_{\mu\to -\mu}.
\ee
At finite chemical potential, $\bra U\ket$ and $\bra U^{-1} \ket$ are both 
real, but different. 
\item
Plaquette:
\be
 \bra \cos x \ket =  \frac{\partial}{\partial\beta}\ln Z = 
 \frac{1}{Z}\left[ I_1(\beta) + 
 \kappa I_1'(\beta) \cosh\mu  \right].
\ee
Note that $\bra\cos x\ket = \half \bra e^{ix} + e^{-ix}\ket$.
\item
Density:
\be
\bra n \ket = \frac{\partial}{\partial\mu} \ln Z = 
\left\bra \frac{i\kappa\sin(x-i\mu)}{1+\kappa\cos(x-i\mu)} \right\ket
= \frac{1}{Z} \kappa I_1(\beta) \sinh\mu.
\ee
At small chemical potential $\bra n \ket$ increases linearly with $\mu$, 
while at large chemical potential $\bra n \ket \to 1$ exponentially fast.
\eit

We now aim to estimate these observables using numerical techniques.
Due to the complexity of the determinant, they cannot be estimated using 
methods based on importance sampling. Instead, we attempt to obtain 
expectation values using stochastic quantization.

At nonzero chemical potential, the action is complex and it becomes 
necessary to consider complex Langevin dynamics. We write therefore 
$x\to z=x+iy$, and consider the following complex Langevin equations
\bea
 x_{n+1} =&&\hm x_n + \eps K_x(x_n,y_n) + \sqrt{\eps}\eta_n, 
\\
 y_{n+1} =&&\hm y_n + \eps K_y(x_n,y_n).
\eea
Here we have discretized Langevin time as $\theta=n\eps$, and the noise 
satisfies
\be 
\bra \eta_n\ket = 0, \;\;\;\;\;\;\;\;
\bra \eta_n \eta_{n'}\ket = 2\delta_{nn'}.
\ee
The drift terms are given by
\be
K_x = -\re \frac{\partial S_{\rm eff}}{\partial x}\Bigg|_{x\to x+iy}, 
\;\;\;\;\;\;\;\;
K_y = -\im \frac{\partial S_{\rm eff}}{\partial x}\Bigg|_{x\to x+iy},
\ee
where the effective action reads
\be
 S_{\rm eff} = S_B - \ln\det M = -\beta\cos x - \ln \left[ 
 1+\kappa\cos(x-i\mu)\right].
\ee
Explicitly, the drift terms are
\bea
K_x =&&\hm  -\sin x\left[\beta\cosh y+\kappa\frac{\cosh(y-\mu)+\kappa\cos 
x}{D(x)}\right],
\\
K_y =&&\hm -\beta\cos x\sinh y    
- \kappa\sinh(y-\mu)\frac{ \cos x+\kappa\cosh(y-\mu)}{D(x)},
\eea
where
\be
D(x) = \left[1+\kappa\cos x\cosh(y-\mu)\right]^2 + \left[\kappa\sin x 
\sinh(y-\mu) \right]^2.
\ee
 Occasionally we will also consider this model by expanding in small 
$\kappa$, the hopping expansion, and take
\be
 \label{eq:hopping}
 S_{\rm eff} = -\beta\cos x - \kappa\cos(x-i\mu)
 \;\;\;\; \;\;\;\;
\mbox{(hopping expansion)}.
\ee
This limit is motivated by the model of Heavy Dense Matter used in Ref.\ 
\cite{DePietri:2007ak}. A direct application of our method to QCD in the hopping expansion is presented in Sec.\ \ref{sec:polyakov}.

In order to compute expectation values, also the observables have to be 
complexified. For example, after complexification $x\to z=x+iy$, the 
plaquette reads 
 \be
\cos x \to \cos(x+iy) =  \cos x\cosh y -i \sin x \sinh y.
\ee
 All operators we consider are now complex, with the real (imaginary) part 
being even (odd) under $x\to -x$.

\begin{figure}
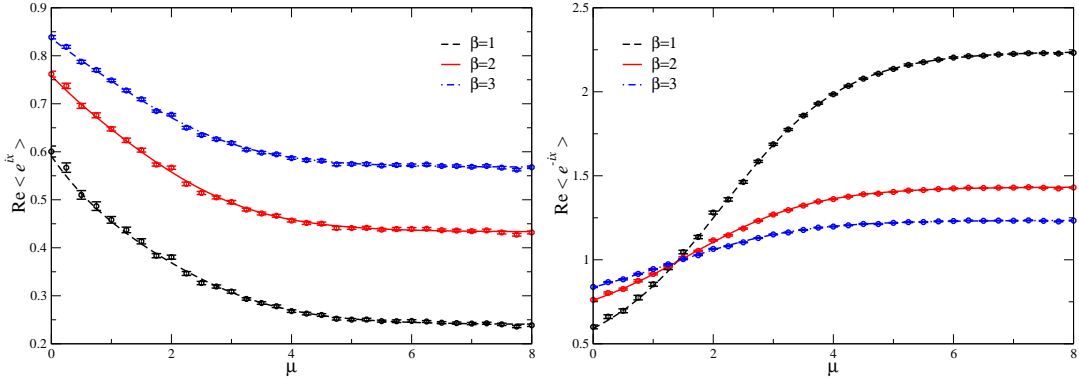

\begin{center}
\epsfig{figure=plot_full_k0.25_poly.eps,height=5cm}
\epsfig{figure=plot_full_k0.25_polycc.eps,height=5cm}
\end{center}
 \caption{Real part of the Polyakov loop $\bra e^{ix}\ket$ (left) 
and the conjugate Polyakov loop $\bra e^{-ix}\ket$ (right) as a 
function of $\mu$ for three values of $\beta$ at fixed $\kappa=1/2$. The 
lines are the analytical results, the symbols are obtained with complex 
Langevin dynamics.}
 \label{fig:poly}
\end{figure}

\begin{figure}
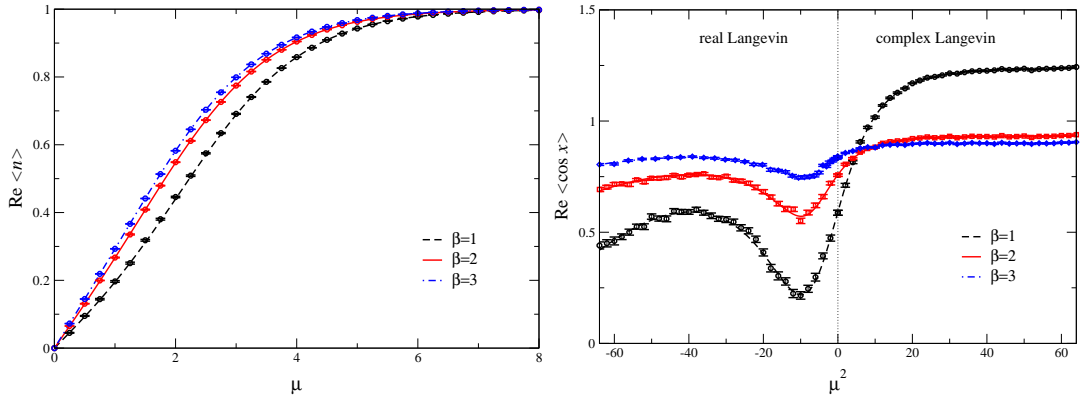
 
\begin{center} 
\epsfig{figure=plot_full_k0.5_density.eps,height=5.2cm} 
\epsfig{figure=plot_full_b1_k0.5_mu2.eps,height=5.2cm} 
\end{center}  
 \caption{Left: Real part of the density $\bra n\ket$.
Right: Real part of the plaquette $\bra \cos x\ket$ versus $\mu^2$. 
Results at positive (negative) $\mu^2$ have been obtained with complex 
(real) Langevin evolution.}
\label{fig:density}
\end{figure}

The Langevin dynamics can be solved numerically. In Fig.\ \ref{fig:poly} 
the real parts of the Polyakov loop and the conjugate Polyakov loop are 
shown as a function of $\mu$ for three values of $\beta$ at fixed 
$\kappa=1/2$. In Fig.\ \ref{fig:density} (left) the density is shown. The 
lines are the exact analytical results. The symbols are obtained with the 
stochastic quantization. We observe excellent agreement between the 
analytical and numerical results. For the results shown here and below, we 
have used Langevin stepsize $\epsilon=5\times 10^{-5}$ and $5\times 
10^{7}$ time steps. Errors are estimated with the jackknife procedure. The 
imaginary part of all observables shown here is consistent with zero 
within the error in the Langevin dynamics.\footnote{Analytically they are 
identically zero.} This can be understood from the symmetries of the drift 
terms and the complexified operators, since the drift terms behave under 
$x\to -x$ as
 \be
 K_x(-x,y;\mu) = -K_x(x,y;\mu), \;\;\;\;\;\;\;\; 
 K_y(-x,y;\mu) = K_y(x,y;\mu), 
\ee 
 while the imaginary parts are odd. Therefore, after averaging over the 
Langevin trajectory the expectation value is expected to reach zero within 
the error, which is what we observe.
 As an aside, we note that the symmetry of the drift terms under $y\to 
-y$,
\be
K_x(x,-y;-\mu) = K_x(x,y;\mu), 
\;\;\;\;\;\;\;\;
K_y(x,-y;-\mu) = -K_y(x,y;\mu),
\ee
relates positive and negative chemical potential.

At imaginary chemical potential $\mu=i\mu_I$ the determinant is 
real, so 
that the complexification of the Langevin dynamics is not necessary. We 
demonstrate the smooth connection for results obtained at imaginary $\mu$ 
using real Langevin dynamics and results obtained at real $\mu$ using 
complex Langevin dynamics for the expectation value of the plaquette 
$\bra\cos x\ket$ in Fig.\ \ref{fig:density} (right). Since the plaquette 
is even under $\mu\to -\mu$, we show the result as a function of $\mu^2$, 
so that the left side of the plot corresponds to imaginary chemical 
potential, while the right side corresponds to real chemical potential. On 
both sides excellent agreement with the analytical expression can be 
observed. We also note that the errors are comparable on both sides.

\subsection{Phase of the determinant} 
\label{sec:det}

At finite chemical potential the determinant is complex and can be written 
as
\be
 \det M = |\det M| e^{i\phi}.
\ee
In order to assess the severity of the sign problem, we consider the phase 
of the determinant and study the behaviour of $e^{i\phi}$. An observable 
often used for this purpose \cite{Splittorff:2006fu,Han:2008xj} is
\be
\label{eq:detdetstar}
 \bra e^{2i\phi} \ket = 
\left\langle \frac{\det M(\mu)}{\det M(\mu)^*} \right\rangle =
\left\langle \frac{\det M(\mu)}{\det M(-\mu)} \right\rangle,
\ee
where we used Eq.\ (\ref{eqdetsym}).
At zero chemical potential the ratio equals one, while at large $\mu$ one 
finds in this model that  
\be
 \lim_{\mu\to \infty} \bra e^{2i\phi} \ket = \frac{I_3(\beta)}{I_1(\beta)} 
+{\cal O}(e^{-\mu}),
\ee
for nonzero $\beta$.
 In expressing Eq.\ (\ref{eq:detdetstar}) as the expectation value 
obtained from the complex Langevin process, complex conjugation has to be 
performed after the complexification of the variables, as discussed above. 
In that case $\det M(\mu)^*$ as a complex number is not the complex 
conjugate of $\det M(\mu)$. To avoid confusion we write ${\det M(-\mu)}$ 
in all relevant expressions. Notice that this implies that $\phi$ itself 
is also complex.

\begin{figure}
\begin{center}
\epsfig{figure=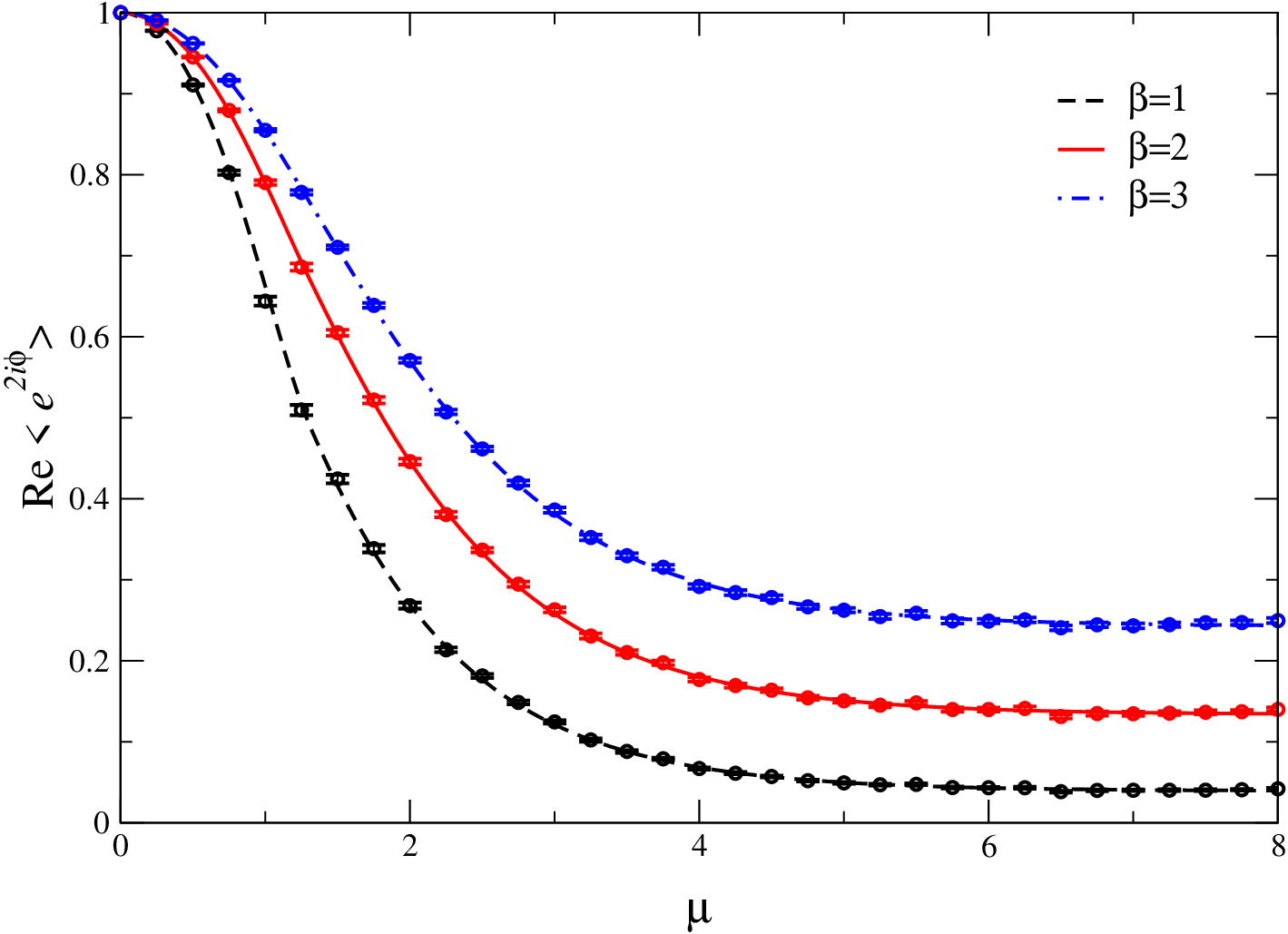,height=5.2cm}
\epsfig{figure=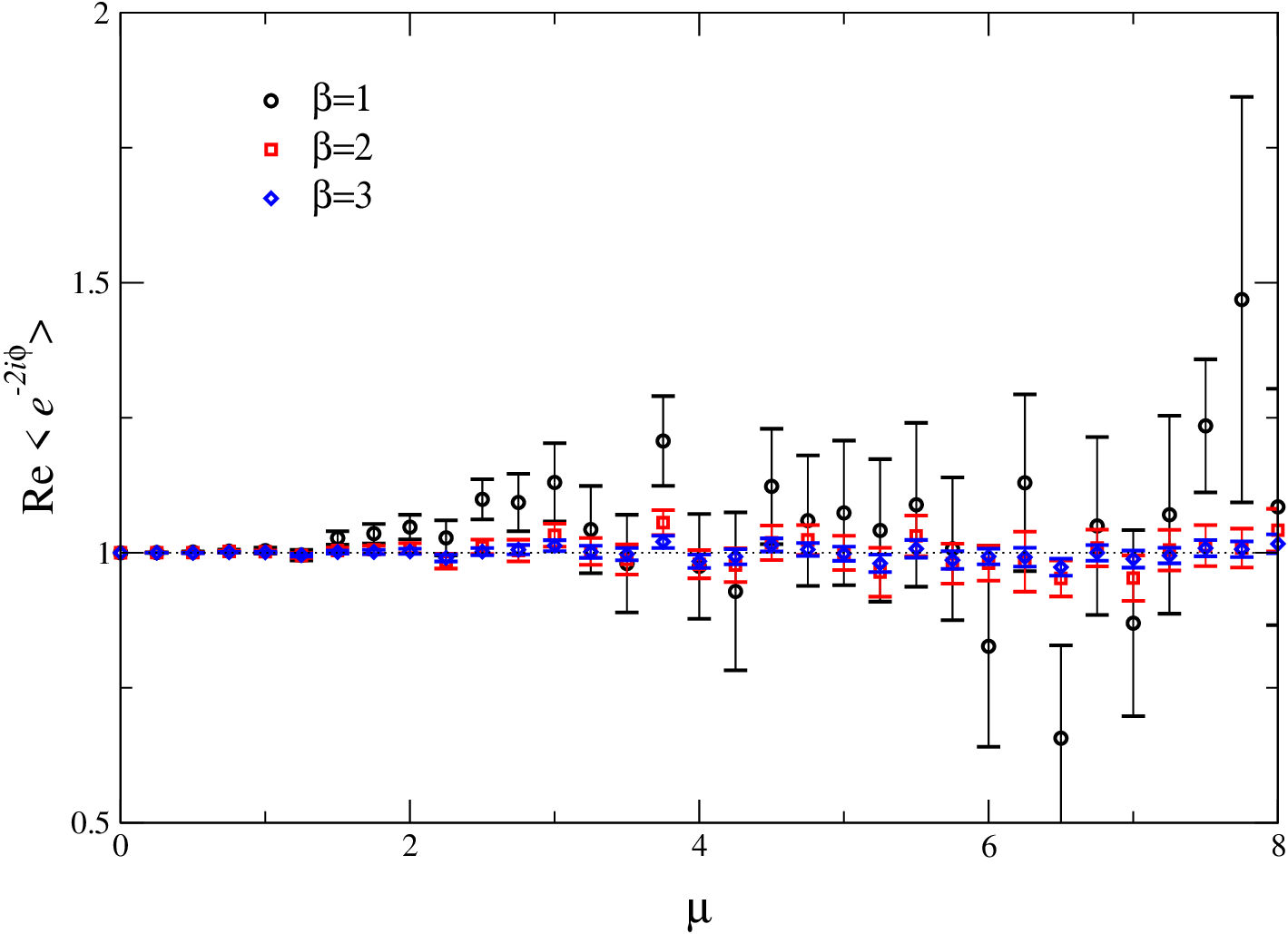,height=5.2cm}
\end{center}
 \caption{Left: Real part of $\bra e^{2i\phi}\ket = \bra \det M(\mu)/\det 
M(-\mu)\ket$. Right: Real part of $\bra e^{-2i\phi}\ket = \bra \det 
M(-\mu)/\det M(\mu)\ket = Z(-\mu)/Z(\mu)$. 
} 
\label{fig:det} \end{figure}

In Fig.\ \ref{fig:det} (left) we show the real part of this observable as 
a function of $\mu$. The imaginary part is again zero analytically and 
zero within the error in the Langevin process. The lines are obtained by 
numerical integration of the observable with the complex weight, while the 
symbols are again obtained from Langevin dynamics. We note again excellent 
agreement between the semi-analytical and the stochastic results. In 
particular, there seems to be no problem in accessing the region with 
larger $\mu$ where the average phase factor becomes very small. The 
numerical error is under control in the entire range.
 We find therefore that the sign problem does not appear to be a problem 
for this method in this model.\footnote{In QCD, the average phase factor 
is expected to go to zero exponentially fast in the thermodynamic limit.}

\begin{figure}[!p]
\begin{center}
\epsfig{figure=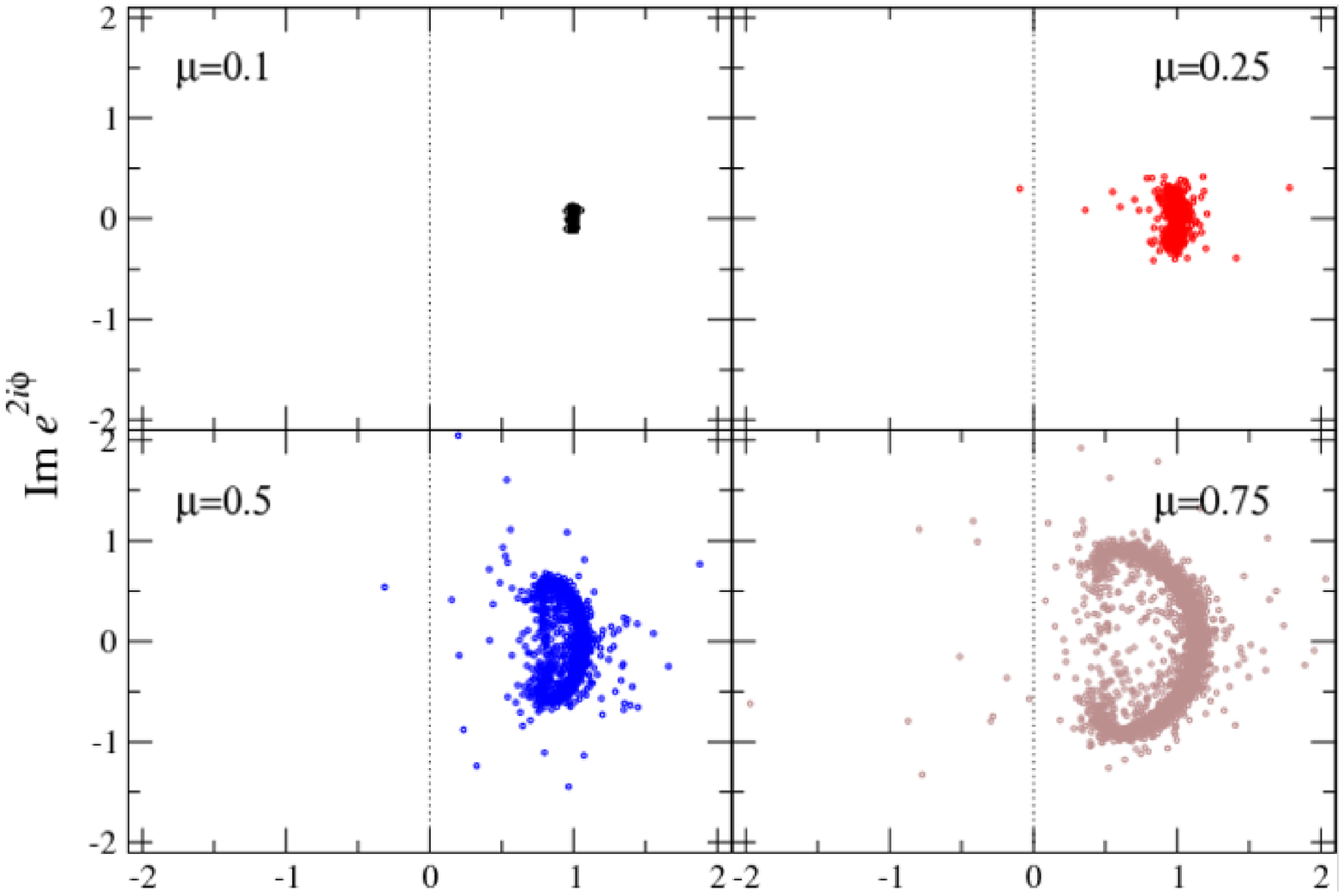,width=9.5cm}
\epsfig{figure=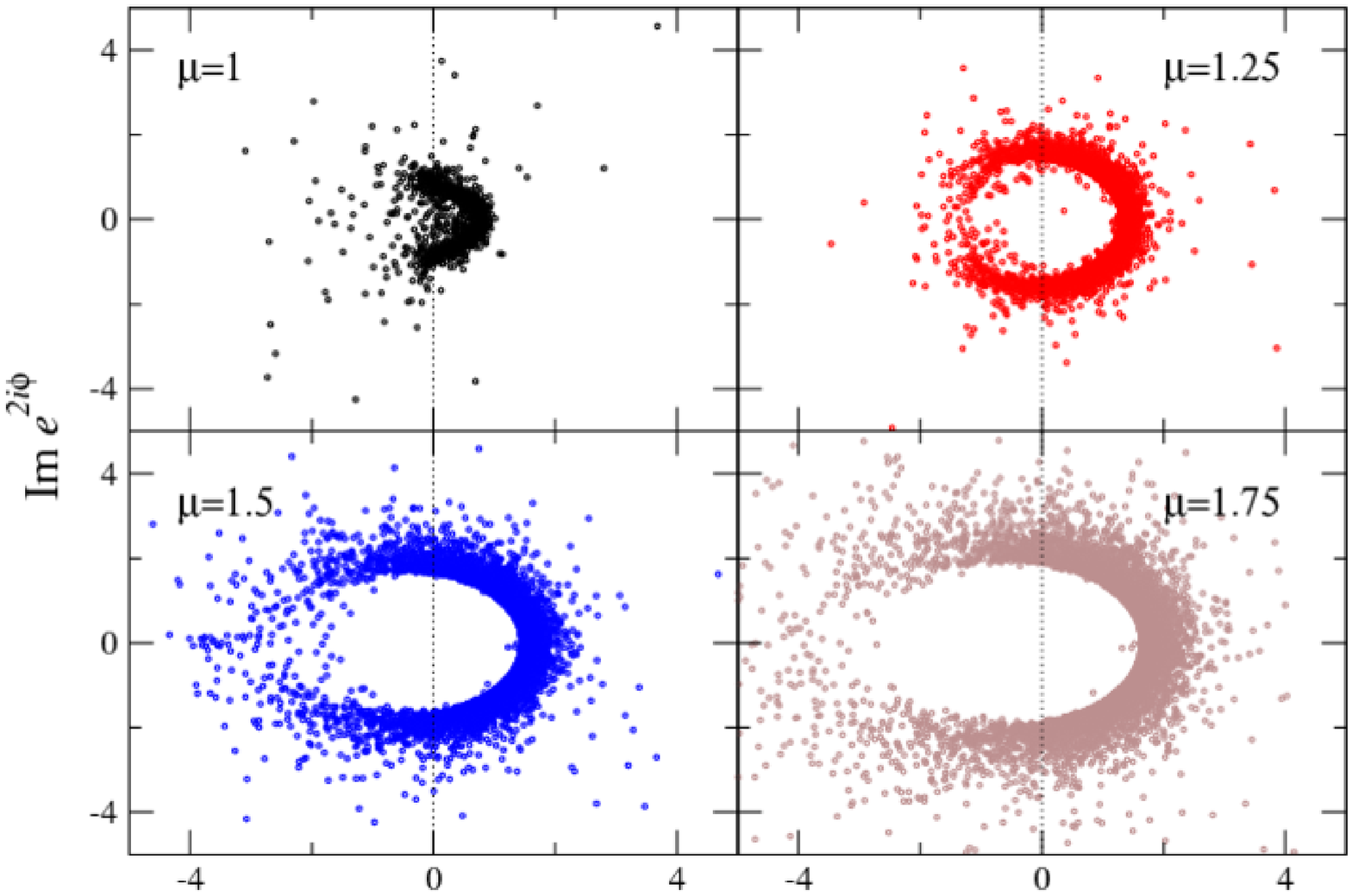,width=9.5cm}
\epsfig{figure=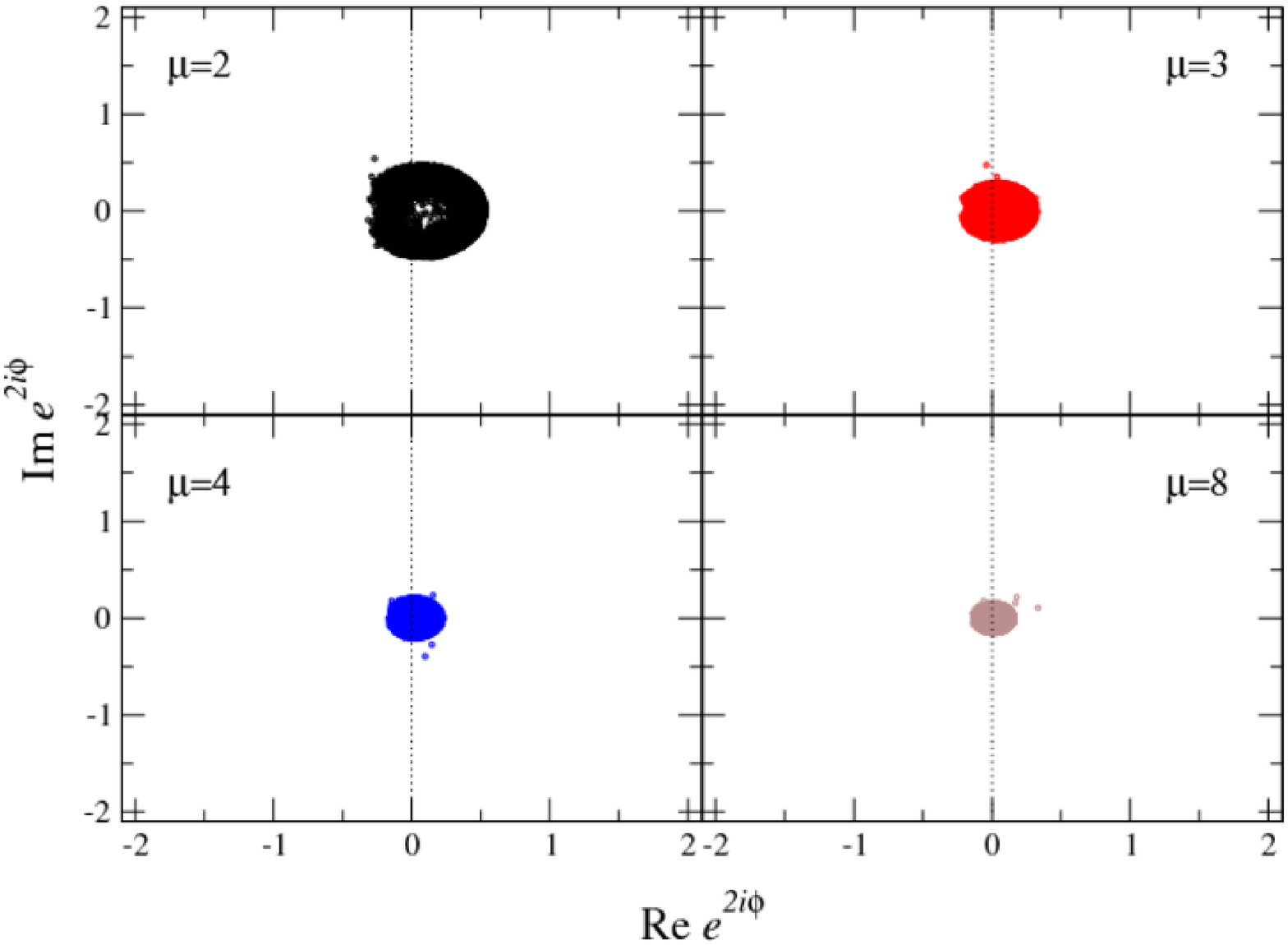,width=9.5cm}
\end{center}
 \caption{Scatter plot of $e^{2i\phi} = \det M(\mu)/\det
M(-\mu)$ during the Langevin evolution for various values of $\mu$ at 
$\beta=1$, $\kappa=1/2$. Note the different scale in the middle box.
}
\label{fig:scat}
\end{figure}

In contrast to what could be inferred from the result above, expectation 
values of $e^{i\phi}$ are not phase factors, due to 
the complexity of the action. This can be seen by considering  
\be
\bra e^{-2i\phi} \ket =  \left\langle \frac{\det M(-\mu)}{\det M(\mu)} 
\right\rangle 
= \frac{Z(-\mu)}{Z(\mu)} = 1, 
\ee
where the second and third equality follow from the cancelation of det 
$M(\mu)$ in the definition of the expectation value and from $Z$ being even 
in $\mu$.
We have also computed this observable using Langevin dynamics and the 
result is shown in Fig.\ \ref{fig:det} (right).  
 For the Langevin parameters used here, we observe that the numerical 
estimate is consistent with 1, but with quite large errors when $\mu$ 
increases at small values of $\beta$. We have found that increasing the 
Langevin time reduces the uncertainty. We conclude that at large chemical 
potential this ratio of determinants is the most sensitive and 
slowest converging observable we considered.

We observed above that the average phase factor becomes very small at 
large $\mu$ but that this does not manifest itself in all but one 
observable we consider. Insight into this feature can be gained by 
studying scatter plots of the phase factor during the Langevin process. In 
Fig.\ \ref{fig:scat} we show the behaviour of $e^{2i\phi}$ during the 
Langevin evolution in the two-dimensional plane spanned by $\re 
e^{2i\phi}$ and $\im e^{2i\phi}$. At zero chemical potential, $\re 
e^{2i\phi}=1$ and $\im e^{2i\phi}=0$ during the entire evolution. For 
increasing $\mu$ one finds more and more deviations from this, with an 
interesting structure at intermediate values of $\mu$. Note that the 
resulting distribution is approximately invariant under reflection in $\im 
e^{2i\phi} \to -\im e^{2i\phi}$, ensuring that the imaginary part of the 
expectation value $\bra e^{2i\phi}\ket$ vanishes within the error. Due to 
the wide distribution, the horizontal and vertical scales in the middle 
section of Fig.\ \ref{fig:scat} are much larger than in the top and bottom 
part. However, the average phase factor remains well defined for all 
values of $\mu$, as can be seen in Fig.\ \ref{fig:det}. At large $\mu$, 
the average phase factor becomes very small. However, the distribution is 
very narrow, see Fig.\ \ref{fig:scat} (bottom). Therefore, although the 
average is close to zero, the error in the Langevin dynamics is very well 
under control.

\subsection{Fixed points and classical flow}

The excellent results obtained above can partly be motivated by the 
structure of the dynamics in the classical limit, i.e.\ in absence of the 
noise. As we demonstrate below, the classical flow and fixed point 
structure is easy to understand when $\mu=0$ and, most importantly, does 
not change qualitatively in the presence of nonzero chemical potential.

Classical fixed points are determined by the extrema of the classical 
(effective) action, i.e.\ by putting $K_x=K_y=0$. We first consider the 
``bosonic'' model and take $\kappa=0$. The drift terms are\footnote{For 
the bosonic model, there is of course no need to complexify the Langevin 
dynamics and one may take $y=0$. This yields the same fixed points.}
\be
K_x(x,y)=-\beta\sin x\cosh y, 
\;\;\;\;\;\;\;\;
K_y(x,y)=-\beta\cos x\sinh y.
\ee
We see that there is one stable fixed point at $(x,y) = (0,0)$ and one 
unstable fixed point at $(\pi,0)$. Moreover, the classical flow equation, 
$dz/d\theta=-\beta \sin z$, can be solved analytically, with the solution
\be
\tan \frac{z(\theta)}{2}= e^{-\beta(\theta-\theta_0)} \tan\frac{z(0)}{2},
\ee
where $z(0)$ is the initial value at $\theta=\theta_0$.
We find therefore that the stable fixed point is reached for all $z(0)$, 
except when $x(0)=\pi$. On this line the solution reads
\be
\tanh \frac{y(\theta)}{2}= e^{\beta(\theta-\theta_0)} \tanh\frac{y(0)}{2},
\ee
and the flow diverges to $y\to \pm\infty$, except when starting 
precisely on the unstable fixed point $(\pi,0)$. Note, 
however, that the noise in the $x$ direction will kick the dynamics of 
the unstable trajectories.

We now include the determinant, starting with the hopping expansion 
(\ref{eq:hopping}). Putting $K_x=K_y=0$ yields again one stable fixed 
point at $(x,y)=(0,y_*)$ and one unstable fixed point at $(\pi,y_*)$, 
where 
 \be
 \tanh y_* = \frac{\kappa\sinh\mu}{\beta+\kappa\cosh\mu}.
 \ee
 Note that in the strong coupling limit $y_*=\mu$. We find therefore a 
simple modification of the bosonic model: in response to the chemical 
potential the two fixed points move in the vertical $y$ direction, but 
not in the $x$ direction.

We continue with the full determinant included. Consider the case 
with $\mu=0$ first, where real dynamics can be considered. Again we find 
the stable fixed point at 
$x=0$ and the unstable fixed point at $x=\pi$. Provided that
 \be 
 \gamma\equiv \left(\frac{1}{\beta}+\frac{1}{\kappa} \right) > 1,
 \ee
 there are no additional fixed points. In order to satisfy this condition, 
we take $\kappa<1$ throughout. Using complex dynamics, while keeping 
$\mu=0$, we find that the stable fixed point at $(x,y)=(0,0)$ remains, but 
that there are now three unstable fixed points at $x=\pi$, given by 
$(\pi,0)$ and $(\pi, \pm y_*)$, with $\cosh y_* = \gamma$. Interestingly, the 
fixed-point structure is therefore different for real and complex flow.

\begin{figure}[!p]
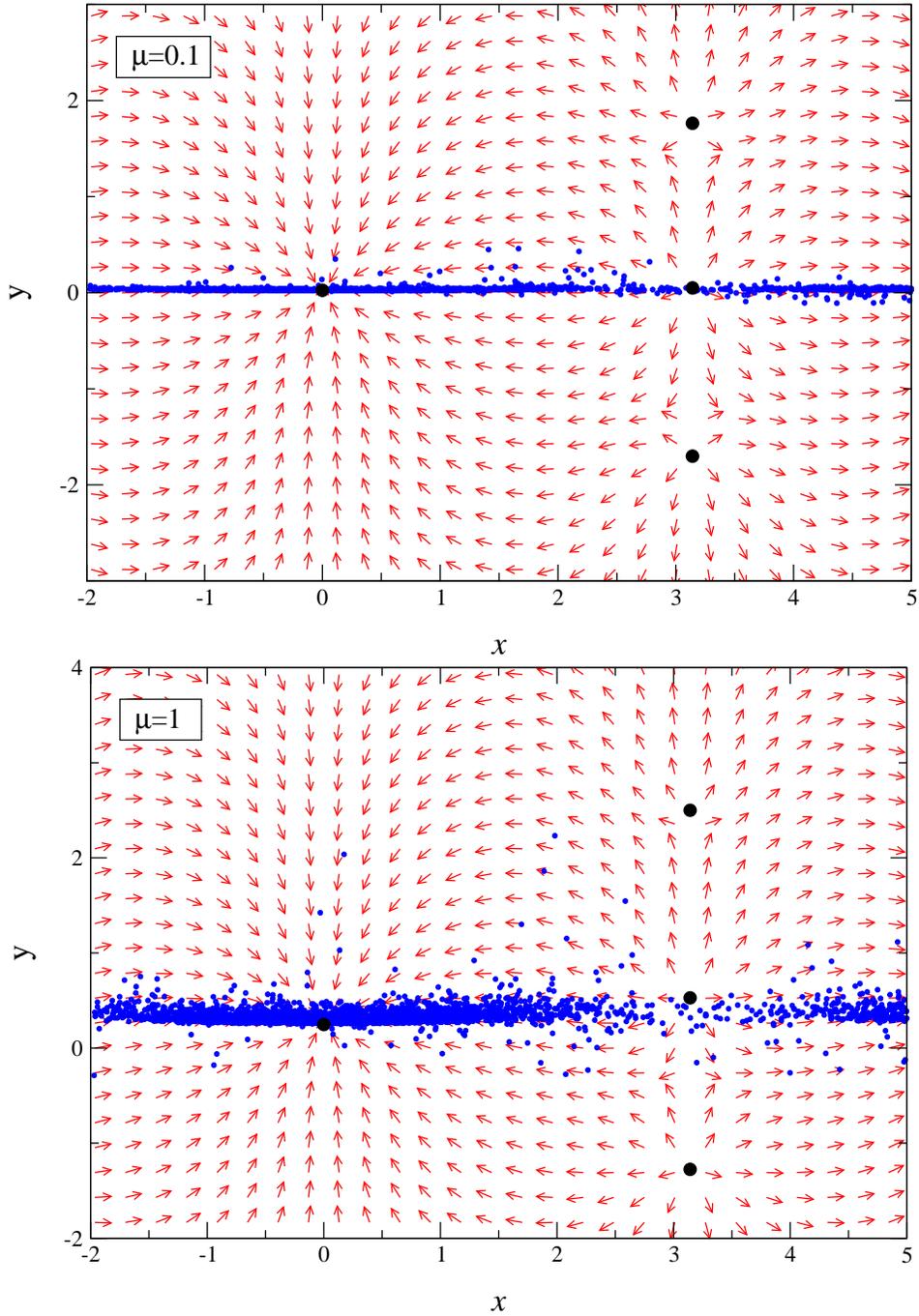

\begin{center}
\epsfig{figure=plot_full_b1_k0.25_m0.1_scat.eps,height=9cm}
\epsfig{figure=plot_full_b1_k0.25_m1_scat.eps,height=9cm}
\end{center}
 \caption{Classical flow diagram in the $x-y$ plane for $\beta=1$, 
$\kappa=1/2$, $\mu=0.1$ (top) and $\mu=1$ (bottom).
The big dots indicate the fixed points at 
$x=0$ and $\pi$. The small circles indicate a trajectory during the 
Langevin evolution, each dot separated from the previous one by 500 
steps. Note the periodicity $x\to x+2\pi$. }
 \label{fig:flow_full_1}
\end{figure}

\begin{figure}[!p]
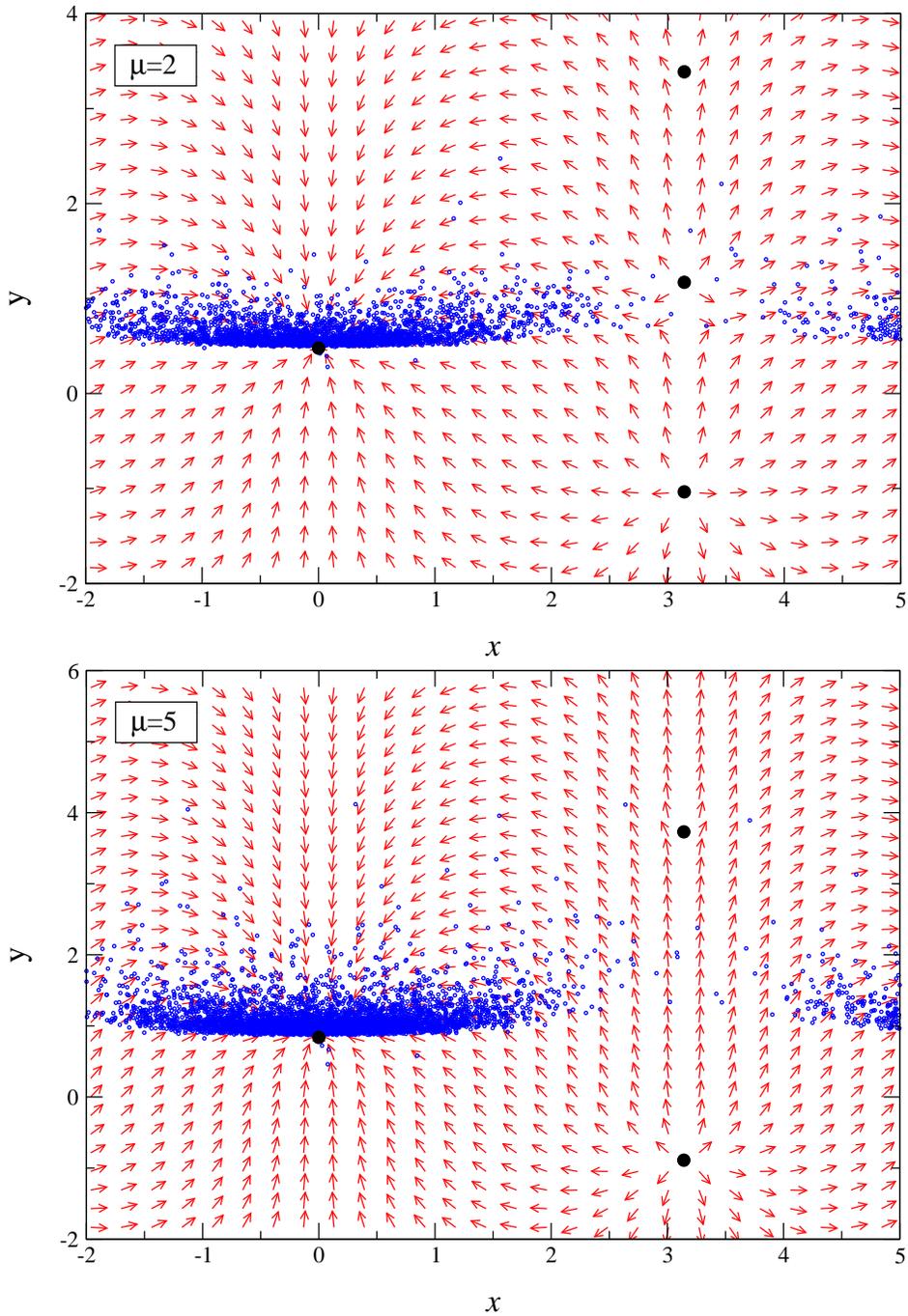

\begin{center}
\epsfig{figure=plot_full_b1_k0.25_m2_scat.eps,height=9cm}
\epsfig{figure=plot_full_b1_k0.25_m5_scat.eps,height=9cm}
\end{center}
 \caption{As in the previous figure, with $\mu=2$ (top) and $\mu=5$ 
(bottom). 
}
 \label{fig:flow_full_2}
\end{figure}

Finally, we come to the full determinant at finite chemical potential. 
In this case the fixed points can only be determined numerically. We find 
a stable fixed point at $(x,y)=(0,y_s)$ and unstable 
fixed points at $(x,y)=(\pi,y_u)$. The $y$ coordinates of these fixed 
points are determined by
 \be
 K_y \Big|_{x=0,\pi} = \mp \beta\sinh y \mp
\frac{\kappa\sinh(y-\mu)}{1\pm\kappa\cosh(y-\mu)} = 0,
 \ee
 where the upper (lower) sign applies to $y_s$ ($y_u$). At $x=0$ there is 
only one solution, while at $x=\pi$ we find numerically that there are 
three (unstable) solutions for small chemical potential, two for 
intermediate $\mu$ and only one for large $\mu$. Although the number of 
fixed points at $x=\pi$ depends on $\mu$, we find that they are always 
unstable such that this has no effect on the dynamics, which is attracted 
to the stable fixed point at $x=0$.

In Figs.\ \ref{fig:flow_full_1} and \ref{fig:flow_full_2} we show the 
classical flow diagrams in the $x-y$ plane. The direction of the arrows 
indicates $(K_x, K_y)$, evaluated at $(x,y)$. The lengths of the arrows 
are normalized for clarity. The fixed points are indicated with the larger 
black dots. In the bosonic model ($\kappa=0$), the analytical solution 
showed that the fixed point at $x=0$ is globally attractive, except when 
$x=\pi$. At nonzero $\kappa$ and $\mu$, the fixed point at $x=0$ appears 
to be globally attractive as well, except again for $x=\pi$. The small 
(blue) dots are part of a Langevin trajectory, each dot separated from the 
previous one by 500 steps. For vanishing $\mu$, the dynamics takes place 
in the $x$ direction only; for increasing $\mu$ it spreads more and more 
in the $y$ direction. An interesting asymmetry around the classical fixed 
point in the $y$ direction can be observed. However, the dynamics remains 
well contained in the $x-y$ plane.

We conclude therefore that the complex Langevin dynamics does not change 
qualitatively in the presence of a chemical potential, small or large. We 
take this as a strong indication that the method is insensitive to the 
sign problem.

\subsection{Fokker-Planck equation}

\label{sec:fpeq}

The microscopic dynamics of the Langevin equation,
\be
\frac{\partial x}{\partial\theta} = -\frac{\partial S}{\partial x} + \eta,
\ee
where $\theta$ is the (continuous) Langevin time, 
can be translated into the dynamics of a distribution $P(x,\theta)$, 
via the relation
 \be
 \bra O(x,\theta) \ket_\eta = \int \frac{dx}{2\pi} P(x,\theta) O(x).
 \ee
From the Langevin equation, it follows that  $P(x,\theta)$ satisfies 
a Fokker-Planck equation,
\be
 \label{eqFP}
 \frac{\partial}{\partial \theta} P(x,\theta) = L_{\rm FP}^c P(x,\theta),
\ee
where $L_{\rm FP}^c$ is the complex Fokker-Planck operator
\be
 L_{\rm FP}^c = 
\frac{\partial}{\partial x} \left(
 \frac{\partial}{\partial x} + \frac{\partial S}{\partial x} \right).
\ee
The stationary solution of the Fokker-Planck equation is easily found by 
putting $L_{\rm FP}^c P(x)=0$ and is given by $P_{\rm st}(x) \sim 
\exp[-S(x)]$. 
In order to cast Eq.\ (\ref{eqFP}) as an eigenvalue
problem, we write $P(x,\theta)=e^{-\lambda\theta}P^{(\lambda)}(x)$.
The solution of the Fokker-Planck equation can then be written as
\be
\label{eq:FPsol}
 P(x,\theta) = \frac{e^{-S(x)}}{Z} + \sum_\lambda 
e^{-\lambda\theta}P^{(\lambda)}(x).
\ee
It is therefore interesting to study the properties of the
Fokker-Planck equation and the nonzero eigenvalues $\lambda$.

\begin{figure}[t]
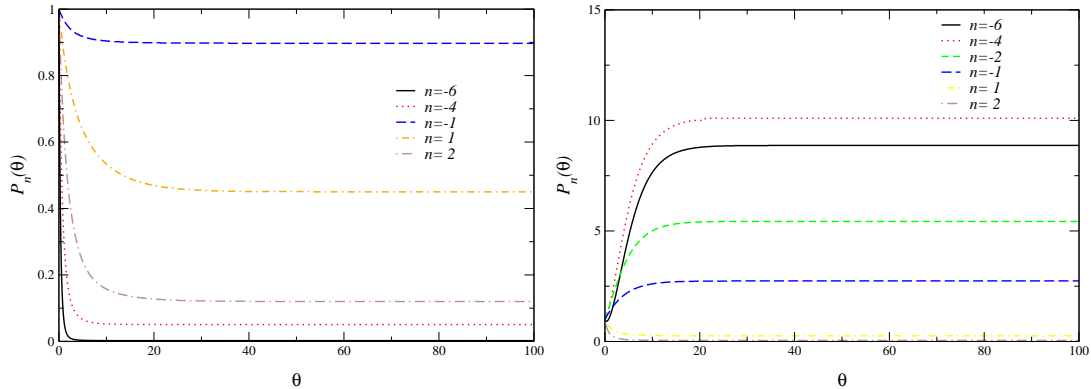

\begin{center}
\epsfig{figure=fp_b1_k0.5_mu1.eps,height=5.1cm}
\epsfig{figure=fp_b1_k0.5_mu3.eps,height=5.1cm}
\end{center}
 \caption{Solution of the complex Fokker-Planck equation: Langevin time 
dependence of the modes $P_n(\theta)$ for various values of $n$, with
$\beta=1$, $\kappa=1/2$, and $\mu=1$ (left) and $\mu=3$ (right).
}
 \label{fig:fptheta}
\end{figure}

In order to do so, we consider the model in the hopping expansion
(\ref{eq:hopping}), with the action
\be
 S = - \beta\cos x - \kappa \cos(x-i\mu).
\ee
 Explicitly, the Fokker-Planck equation then reads
\bea
\dot P(x,\theta) =&&\hm
P''(x,\theta) +
\left[ \beta\sin x + \kappa \sin(x - i\mu)\right] P'(x,\theta) 
\nn \\
&&\hm +\left[ \beta\cos x + \kappa \cos(x - i\mu)\right] P(x,\theta), 
\eea
where primes/dots indicate $x$/$\theta$ derivatives.
  Using periodicity, $P(x+2\pi,\theta) =  P(x,\theta)$, we decompose
\be
 P(x,\theta) =  \sum_{n\in \mathbb{Z}} e^{-inx} P_n(\theta), \\
\;\;\;\;\;\;\;\;
 P_n(\theta) =  \int_{-\pi}^\pi \frac{dx}{2\pi}\, e^{inx} P(x,\theta),
\ee
 and we find
\be
 \label{eq:FPtheta}
\dot P_n(\theta) =
-n^2P_n(\theta) - n c_+P_{n+1}(\theta) + n c_-P_{n-1}(\theta),
\ee
with
\be
 c_\pm = \half\left(\beta+\kappa e^{\pm \mu}\right).
\ee
 We note that this equation is completely real, such that all 
$P_{n}(\theta)$'s are real. This is expected for the stationary solution, 
since from $S^*(x)=S(-x)$ it follows that $P^*_{\rm st}(x)=P_{\rm 
st}(-x)$ and therefore $P_{n, \rm st}^* = P_{n, \rm st}$. The numerical 
solution of Eq.\ (\ref{eq:FPtheta}) is shown in Fig.\ \ref{fig:fptheta} 
for a number of modes $P_n(\theta)$ for $\mu=1$ (left) and 3 (right). The 
initial values $P_n(0)=1$ for all $n$. The number of modes is truncated, 
with $-50<n<50$. For large $\pm n$, $P_n(\theta)\to 0$ exponentially 
fast. The zero mode $P_0$ is $\theta$ independent and equal to 1 by 
construction. We have verified that the other modes converge to the 
values determined by the stationary solution $\sim e^{-S}$.

The convergence properties can be understood from the eigenvalues of the 
Fokker-Planck operator. Writing $P_n(\theta) = \exp(-\lambda\theta)P_n$ 
gives the eigenvalue equation
 \be
 \label{eq:FP}
 n^2P_n + n c_+ P_{n+1} - n c_- P_{n-1} = \lambda P_n.
\ee
 Since all $P_n$ are real, also all eigenvalues $\lambda$ are real. 
 Although at first sight this may seem surprising, it is a consequence of 
the symmetry of the action and therefore it also holds in e.g.\ the full 
model.

\begin{figure}[t]
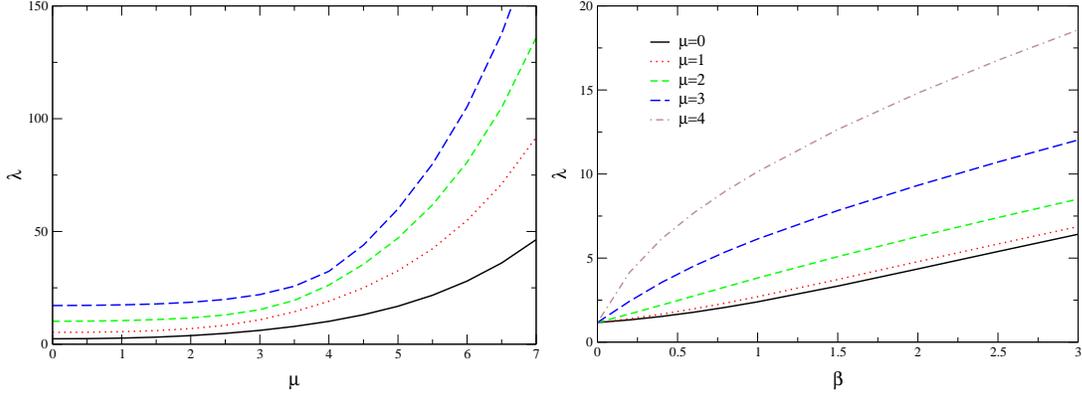

\begin{center}
\epsfig{figure=fokker_eigenvalues_b1_k0.5.eps,height=5.2cm}
\epsfig{figure=fokker_eigenvalues_k0.5_mu0-4.eps,height=5.2cm}
\end{center}
 \caption{Left: Four smallest nonzero eigenvalues of the 
complex Fokker-Planck 
equation as a function of $\mu$ with $\beta=1$, $\kappa=1/2$.
 Right: Smallest nonzero eigenvalue as a 
function of $\beta$ for various values of $\mu$ at $\kappa=1/2$.
}
 \label{fig:fp}
\end{figure}

The case $\lambda=0$ corresponds to the stationary solution. Here we 
consider $\lambda\neq 0$. First take $n=0$. It follows 
from Eq.\ (\ref{eq:FP}) that $P_0=0$. As a result, the sequences for $n>0$ 
and $n<0$ split in two. Written in matrix form, they read
 \be
 \left(
 \begin{array}{ccccc}
 1 & c_\pm & 0  & 0 &\ldots\\
 -2c_\mp & 4 & 2c_\pm & 0 & \ldots\\
 0 & -3c_\mp & 9 & 3c_\pm & \ldots\\
 0 & 0 & -4c_\mp & 16 & \ldots  \\
 \vdots & \vdots & \vdots & \vdots & \ddots
 \end{array}
 \right) 
  \left(
 \begin{array}{c}
 P_{\pm 1} \\
 P_{\pm 2} \\
 P_{\pm 3} \\
 P_{\pm 4} \\
 \vdots
 \end{array}
 \right)
 =
 \lambda
   \left(
 \begin{array}{c}
 P_{\pm 1} \\
 P_{\pm 2} \\
 P_{\pm 3} \\
 P_{\pm 4} \\
 \vdots
 \end{array}
 \right)
\ee 
 Approximating this matrix by a large but finite matrix, one can easily 
compute the eigenvalues numerically. We find that they are all positive 
and that the $n>0$, $n<0$ sequences yield identical eigenvalues. In Fig.\ 
\ref{fig:fp} (left) the four smallest nonzero eigenvalues are shown as a 
function of chemical potential. All eigenvalues are strictly positive and 
increase with $\mu$.  In Fig.\ \ref{fig:fp} (right) the dependence on 
$\beta$ is indicated. At vanishing $\beta$, the $\mu$ dependence cancels, 
since in that case $c_+c_-=\kappa^2/4$. Also as a function of $\beta$ we 
observe that the eigenvalues are strictly positive.

If the action and therefore the Langevin dynamics would be real, these 
results would be sufficient to explain the convergence of the observables 
towards the correct values as observed above, by employing
Eq.\ (\ref{eq:FPsol}) in the large $\theta$ limit. In the complex case we 
consider here, this is not immediately clear. Given the real Langevin 
equations, 
\be
\frac{\partial x}{\partial\theta} = K_x + \eta,
\;\;\;\;\;\;\;\;\;\;\;\;
\frac{\partial y}{\partial\theta} = K_y, 
\ee
one can also consider the real distribution $\rho(x,y,\theta)$,
satisfying the Fokker-Planck equation
\be
 \frac{\partial}{\partial \theta} \rho(x,y,\theta) = L_{\rm FP} 
\rho(x,y,\theta),
\ee
with the real Fokker-Planck operator
\be
L_{\rm FP} = 
 \frac{\partial}{\partial x} \left(\frac{\partial}{\partial x} - K_x \right) 
- \frac{\partial}{\partial y} K_y.
\ee
After complexification, expectation values should then satisfy
 \be
\bra O(x+iy, \theta) \ket_\eta = \int \frac{dxdy}{2\pi}\, \rho(x,y,\theta) 
O(x+iy).
\ee
 In contrast to the complex distribution $P(x,\theta)$, the distribution 
$\rho(x,y,\theta)$ is real and has the interpretation as a probability 
distribution in the $x-y$ plane.
 The real and complex Fokker-Planck operators can be related, using 
\bea
\frac{\partial}{\partial\theta} \bra O(x+iy, \theta) \ket = &&\hm 
\int \frac{dxdy}{2\pi}\, O(x+iy) L_{\rm FP} \rho(x,y,\theta) \nn \\
= &&\hm 
\int \frac{dxdy}{2\pi}\, O(x+iy) L_{\rm FP}^c \rho(x,y,\theta).
\label{eq:rcFP}
\eea
 Here partial integration for finite $\theta$ is used as well as the 
analytic dependence of $O$ on $z=x+iy$. Eq.\ (\ref{eq:rcFP}) suggests a 
relation between the spectrum of the complex and the real Fokker-Planck 
operator. However, we have not yet been able to show that a stationary 
solution of the real Fokker-Planck equation exists. We hope to come back 
to this issue in the future.

\section{One link SU(3) model}
\setcounter{equation}{0}
\label{sec:su3onelink}

\subsection{Model}

In this section we consider a one link model where the link $U$ is an 
element of SU(3). The partition function reads
\be
Z = \int dU\, e^{-S_B} \det M,
\ee
with the bosonic part of the action\footnote{\label{fn1}Note that for an 
SU(3) matrix, $U^{-1}=U^\dagger$. Nevertheless, we write $U^{-1}$ to allow 
for a straightforward complexification of the Langevin dynamics.}
\be
 S_B = -\frac{\beta}{6}\left(\Tr U + \Tr U^{-1}\right).
\ee
For the fermion matrix we take
\be
 M = 1+\kappa\left( e^\mu\sigma_{+} U + e^{-\mu}\sigma_{-} U^{-1} 
\right),
\ee
with $\sigma_\pm=\half(\id\pm\sigma_3)$. We use the Pauli matrix 
$\sigma_3$ rather than $\gamma$ matrices to avoid factors of 2.
The determinant has the product form
\be
 \det M = \det\left(1+\kappa e^\mu U\right)
 \det\left(1+\kappa e^{-\mu} U^{-1}\right),
\ee
where the remaining determinants on the right-hand side are in colour 
space.
In order to exponentiate the determinant, we use the identity, valid for 
$U\in$ SL($3,\mathbb{C}$),
 \be
 \label{eq:dettr}
 \det\left( 1+ c U\right) = 1 + c \Tr U + c^2\Tr U^{-1} + c^3.
\ee
We find therefore that
\be
 \det M = e^{-S_F}, 
\;\;\;\;\;\;\;\; 
S_F = -\ln {\cal M}^{(q)} - \ln  {\cal M}^{(\bar q)}, 
\ee
 with the quark and anti-quark contributions
\bea
{\cal M}^{(q)} =&&\hm 1+3\kappa e^\mu P + 3\kappa^2 e^{2\mu} P^{-1} + 
\kappa^3e^{3\mu},
\\
{\cal M}^{(\bar q)} =&&\hm 1+ 3\kappa e^{-\mu} P^{-1} + 3\kappa^2 e^{-2\mu}P +
\kappa^3e^{-3\mu}.
\eea
Here we introduced the Polyakov loop $P$ and its ``conjugate'' $P^{-1}$,
\be
 P = \frac{1}{3}\Tr U, 
\;\;\;\;\;\;\;\;\;
 P^{-1} = \frac{1}{3}\Tr U^{-1}.
\ee 
Note that $PP^{-1}\neq 1$.
 At large $\mu$, the anti-quark contribution ${\cal M}^{(\bar q)}\to 1$ 
and no longer contributes. However, the term is crucial to preserve the 
symmetry (\ref{eqdetsym}) and is in particular relevant at imaginary and 
small real $\mu$.

Observables are defined as
\be
\bra O(U)\ket = \frac{1}{Z}\int dU\, e^{-S_B(U)} \det M(U)\, O(U).
\ee
The observables we consider are the Polyakov loop $P$, the 
conjugate Polyakov loop $P^{-1}$ and the density $n$. The latter is 
determined by 
\be
 \bra n\ket = \frac{\partial\ln Z}{\partial\mu}, 
\ee 
and reads
\bea
 n = &&\hm 
\frac{\partial \ln {\cal M}^{(q)} }{\partial \mu}
+
\frac{\partial \ln {\cal M}^{(\bar q)} }{\partial \mu}
\nn \\ 
=&&\hm 
3\frac{\kappa e^\mu P + 2\kappa^2 e^{2\mu} P^{-1} + \kappa^3e^{3\mu} }
{{\cal M}^{(q)}}
- 
3\frac{\kappa e^{-\mu} P^{-1} + 2\kappa^2 e^{-2\mu}P +\kappa^3e^{-3\mu} }
{{\cal M}^{(\bar q)}}. 
\;\;\;\;
\eea
At zero chemical potential, the density vanishes while at large 
$\mu$ the density $n \to 3$, the maximal numbers of (spinless) quarks on 
the  site.

In this model, expectation values can be obtained directly 
by numerical integration, allowing for a comparison 
with the results from stochastic quantization presented below.
Since we only consider observables that depend on the conjugacy class of 
$U$, we only have to integrate over the conjugacy classes. These are 
parametrized by two angles $-\pi<\phi_1, \phi_2\leq \pi$. The reduced 
Haar measure on 
the conjugacy classes $[U]$ reads
\be
 d[U] = \frac{1}{\cal N}
 \sin^2\left[\half(\phi_1-\phi_2)\right]
 \sin^2\left[\half(\phi_1+2\phi_2)\right]
 \sin^2\left[\half(\phi_2+2\phi_1)\right],
\ee where ${\cal N}$ is a normalization constant. The matrix is 
parametrized as
\be
 U = \diag\left( e^{i\phi_1}, e^{i\phi_2}, e^{-i(\phi_1+\phi_2)}\right), 
\ee
such that
\be
 S_B = -\frac{\beta}{3}\left[ \cos(\phi_1) + \cos(\phi_2) +
\cos(\phi_1+\phi_2)\right],
\ee
and
\be
 P = \frac{1}{3}\left[ e^{i\phi_1} + e^{i\phi_2} + e^{-i(\phi_1+\phi_2)} \right],
\;\;\;\;
 P^{-1} = \frac{1}{3}\left[ e^{-i\phi_1} + e^{-i\phi_2} + 
e^{i(\phi_1+\phi_2)} \right].
\ee
 It is now straightforward to compute expectation values by numerical 
integration over $\phi_1$ and $\phi_2$.

\subsection{Complex Langevin dynamics}

In contrast to in the U(1) model, the Langevin dynamics is now defined in 
terms of matrix multiplication. We denote $U(\theta+\eps) = U'$ and 
$U(\theta) = U$, where $\theta$ is again the Langevin 
time and consider the Langevin process, 
\be
\label{eq:URU}
 U' = R\, U, 
\;\;\;\;\;\;\;\;
\;\;\;\;\;\;\;\;
R = \exp \left[ i\lambda_a\left( \eps K_a +\sqrt \eps \eta_a\right) 
\right].
\ee
Here $\lambda_a$ ($a=1,\ldots,8$) are the traceless, hermitian Gell-Mann 
matrices, normalized as $\Tr \lambda_a\lambda_b = 2\delta_{ab}$.
The noise satisfies
\be
 \bra\eta_a\ket=0,
\;\;\;\;\;\;\;\;\;
 \bra\eta_a\eta_b\ket=2\delta_{ab},
\ee
and the drift term reads 
\be
  K_a =  -D_a S_{\rm eff}, 
\;\;\;\;\;\;\;\;\;\;\;\;
 S_{\rm eff} = S_B+S_F.
\ee
Differentiation is defined as
\be
 D_{a} f(U) = \frac{\partial}{\partial\alpha} f\left( e^{i\alpha 
\lambda_a} U\right)\Big|_{\alpha=0}.
\ee
In particular, $D_{a}U = i\lambda_a U$ and $D_{a}U^{-1} = -i U^{-1} 
\lambda_a$.

The explicit expressions for the  drift terms  are
\be
 K_a=K_a^B+K_a^F,
\ee
with
\bea
 K_a^B = -D_a S_B(U) = &&\hm \frac{\beta}{2} \left( D_aP + D_aP^{-1} \right),
\\
 K_a^F = -D_a S_F(U) = &&\hm
3\frac{\kappa e^\mu D_aP + \kappa^2 e^{2\mu} D_aP^{-1}}{ {\cal M}^{(q)}}
+3\frac{\kappa e^{-\mu} D_aP^{-1} + \kappa^2 e^{-2\mu} D_aP}{  {\cal 
M}^{(\bar q)}},
\nn\\
&&
\eea
written in terms of
\be
 D_aP = \frac{i}{3} \Tr\lambda_a U,
\;\;\;\;\;\;\;\;\;\;\;\;
 D_aP^{-1} = -\frac{i}{3} \Tr U^{-1}\lambda_a.
\ee
 Let us first consider the case without chemical potential and take $U\in$ 
SU(3). Then it is easy to see that $K^\dagger_a = K_a$ and therefore 
$R^\dagger R=\id$. Moreover, since the Gell-man matrices are traceless, 
$\det R=1$. Therefore, if $U$ is an element of SU(3), it will remain in 
SU(3) during the Langevin process. The same results are found at finite 
imaginary chemical potential $\mu=i\mu_I$. At nonzero real $\mu$ on the 
other hand, we find that $R^\dagger R\neq \id$, although $\det R=1$ still 
holds. Therefore $U$ will be an element of SL($3,\mathbb{C}$) during the 
Langevin evolution. If $U$ is parametrized as $U=\exp \left(i\lambda_a 
A_a/2\right)$, this implies that the gauge potentials $A_a$ are complex.

\begin{figure}[!p]
\begin{center}
\epsfig{figure=plot_full_su3_k0.5_poly.eps,height=5.2cm}
\epsfig{figure=plot_full_su3_k0.5_polycc.eps,height=5.2cm}
\end{center}
 \caption{Real part of 
the Polyakov loop $\bra P\ket$ (left) and the conjugate Polyakov loop 
$\bra P^{-1}\ket$ (right) as a function of $\mu$ for three values 
of $\beta$ at fixed $\kappa=1/2$. 
}
 \label{fig:polysu3}
\vspace*{0.9cm}
\begin{center}
\epsfig{figure=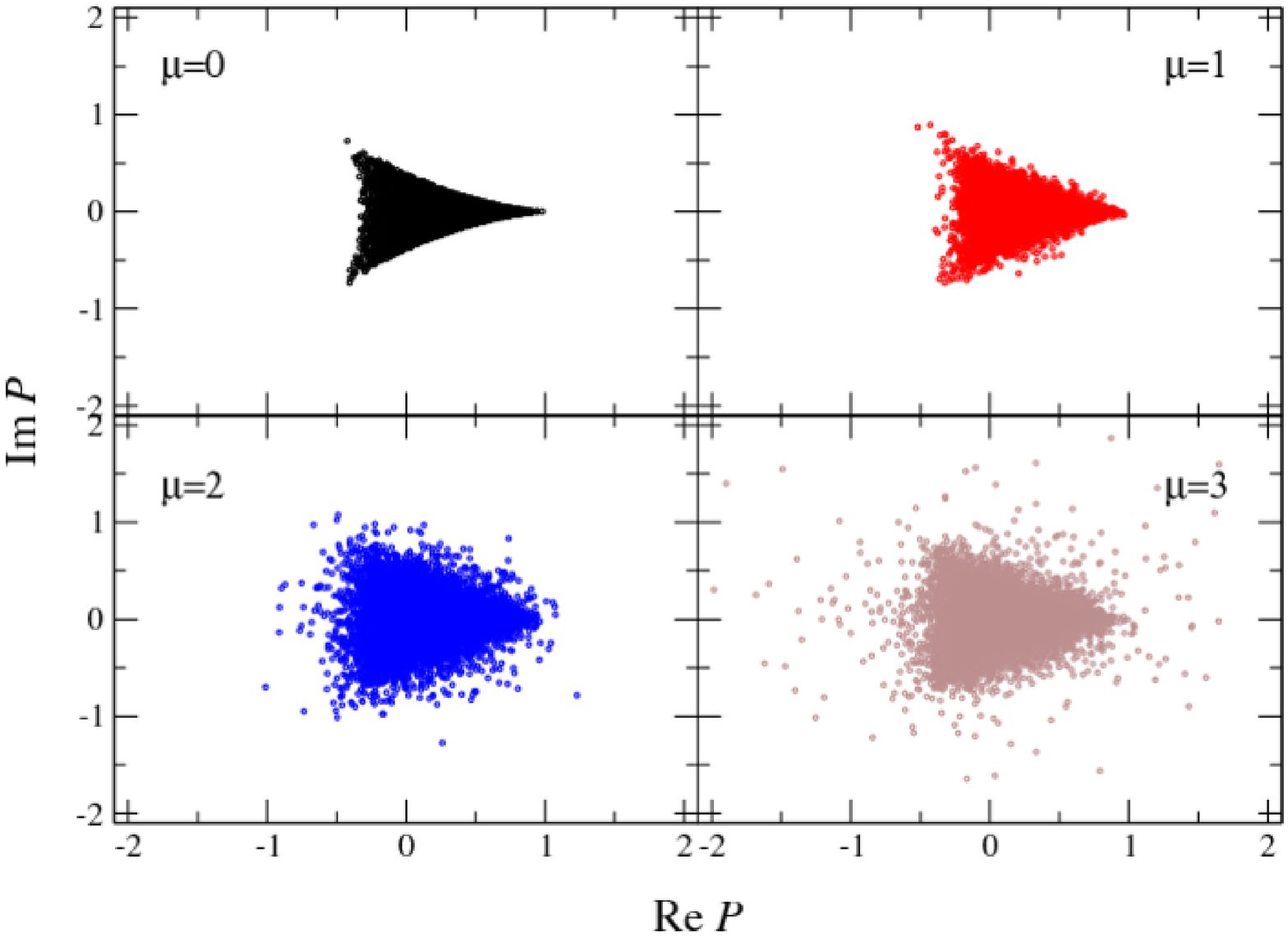,width=12.5cm}
\end{center}
 \caption{Scatter plot of the Polyakov loop for $\beta=1$, $\kappa=1/2$ 
and $\mu=0,1,2,3$.}
\label{fig:scatpoly}
\end{figure}

We solved the Langevin process (\ref{eq:URU}) numerically. The matrix $R$ 
is computed by exponentiating the complex traceless matrix 
$i\lambda_a\left(\eps K_a +\sqrt \eps \eta_a\right)$, employing Cardano's 
method \cite{abra} for finding the eigenvalues. In Fig.\ \ref{fig:polysu3} 
the real part of the Polyakov loop $\bra P\ket$ and the conjugate Polyakov 
loop $\bra P^{-1}\ket$ are shown as a function of $\mu$ for three values of 
$\beta$ at fixed $\kappa=1/2$. The lines are the `exact' results obtained 
by numerically integrating over the angles $\phi_1$ and $\phi_2$, as 
discussed above. The symbols are obtained with complex Langevin dynamics, 
using the same Langevin stepsize and number of time steps as in the U(1) 
model ($\epsilon=5\times 10^{-5}$ and $5\times 10^7$ time steps). Errors 
are estimated with the jackknife procedure. Again, the imaginary part is 
zero analytically and consistent with zero within the error in the Langevin 
dynamics. Excellent agreement between the exact and the stochastic results 
can be seen.

\begin{figure}[t]
\begin{center}
\epsfig{figure=plot_full_su3_k0.5_density.eps,height=5.15cm}
\epsfig{figure=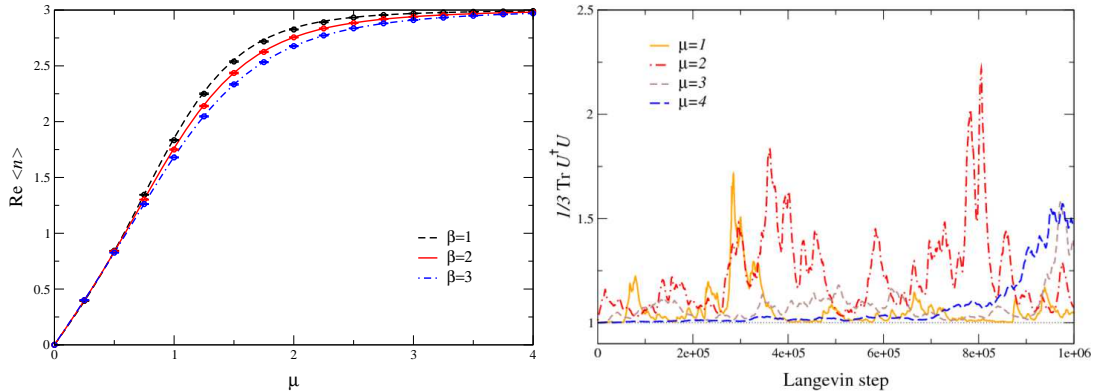,height=5.15cm}
\end{center}
 \caption{Left: Real part of the density $\bra n\ket$.
 Right: Deviation from SU(3) during complex Langevin evolution: $\Tr 
U^\dagger U/3$ as a function of Langevin step, for $\mu=1,2,3,4$ at 
$\beta=1, \kappa=1/2$.
}
\label{fig:densitysu3}
\end{figure}

Scatter plots of the Polyakov loop during the Langevin evolution are shown 
in Fig.\ \ref{fig:scatpoly} for four values of $\mu$ at $\beta=1$ and 
$\kappa=1/2$. Every point is separated from the previous one by 500 time 
steps. Note that the distribution is approximately symmetric under 
reflection $\im P\to -\im P$, ensuring that $\im\bra P\ket=0$ within the 
error. We observe that the characteristic shape visible at $\mu=0$ becomes 
more and more fuzzy at larger $\mu$, but the average remains well defined 
for all values of $\mu$.
 The density is shown in Fig.\ \ref{fig:densitysu3} (left), with again 
good agreement between the exact and the stochastic results. We observe 
that saturation effects set in for smaller values of $\mu$ compared to the 
U(1) model, e.g.\ $n=n_{\rm max}/2=3/2$ already at $\mu\approx 1$.

During the complex Langevin evolution the matrix $U\in$ 
SL$(3,\mathbb{C})$. In order to quantify how much it deviates from SU(3), 
we may follow the evolution of 
\be
 f(U) = \frac{1}{3}\Tr U^\dagger U. \ee It is easy to show that $f(U)\geq 
1$, with the equality in the case that $U\in$ SU(3).\footnote{Consider 
$U\in$ SL$(N,\mathbb{C})$ and $f(U) = \Tr U^\dagger U/N$. Using a polar 
decomposition, $U = VP$, with $V\in$ SU$(N)$ and $P$ a positive 
semidefinite hermitian matrix with $\det P=1$, it is easy to show that 
$f(U)\geq 1$, with the equal sign holding when $U\in$ SU$(N)$.} It 
provides therefore a good measure to quantify the deviation from SU(3). In 
Fig.\ \ref{fig:densitysu3} (right), we show this quantity during the 
Langevin evolution. We observe that the deviations from 1 are present but 
not too large. If $U$ is parametrized as $U=\exp \left(i\lambda_a 
A_a/2\right)$, this implies the imaginary parts of the gauge potentials 
$A_a$ do not become unbounded.

\subsection{Phase of the determinant}
\label{sec:detsu3}

\begin{figure}[t]
\begin{center}
\epsfig{figure=plot_full_su3_k0.5_detdetstar.eps,height=5.1cm}
\epsfig{figure=plot_full_su3_k0.5_detstardet.eps,height=5.1cm}
\end{center}
 \caption{Left: Real part of $\bra e^{2i\phi} \ket = \bra \det M(\mu)/\det 
M(-\mu)\ket$. Right: Real part of $\bra \det M(-\mu)/\det M(\mu)\ket$. }
\label{fig:detdetstarsu3}
\end{figure}

\begin{figure}[!p]
\begin{center}
\epsfig{figure=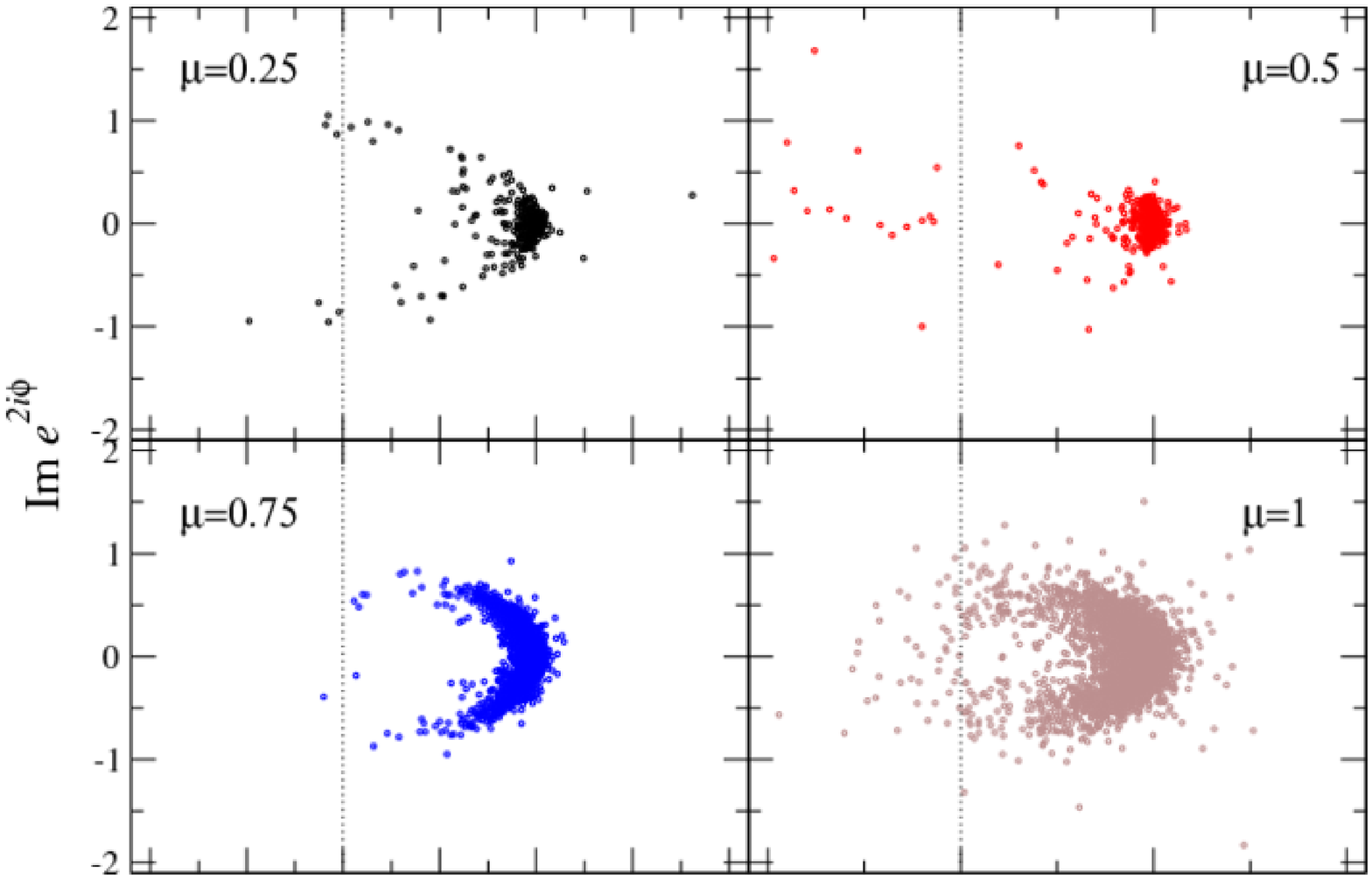,width=9.5cm}
\epsfig{figure=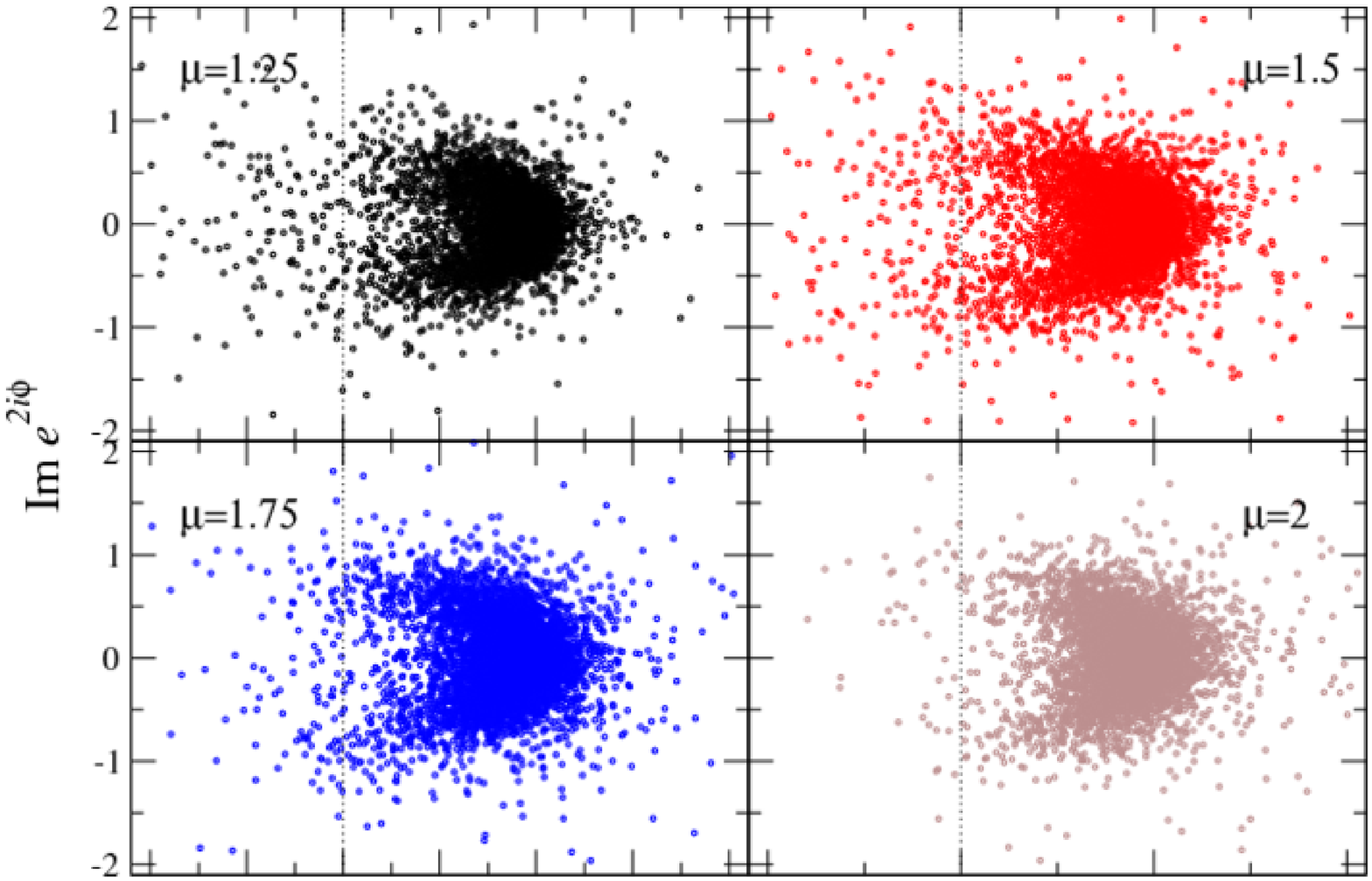,width=9.5cm}
\epsfig{figure=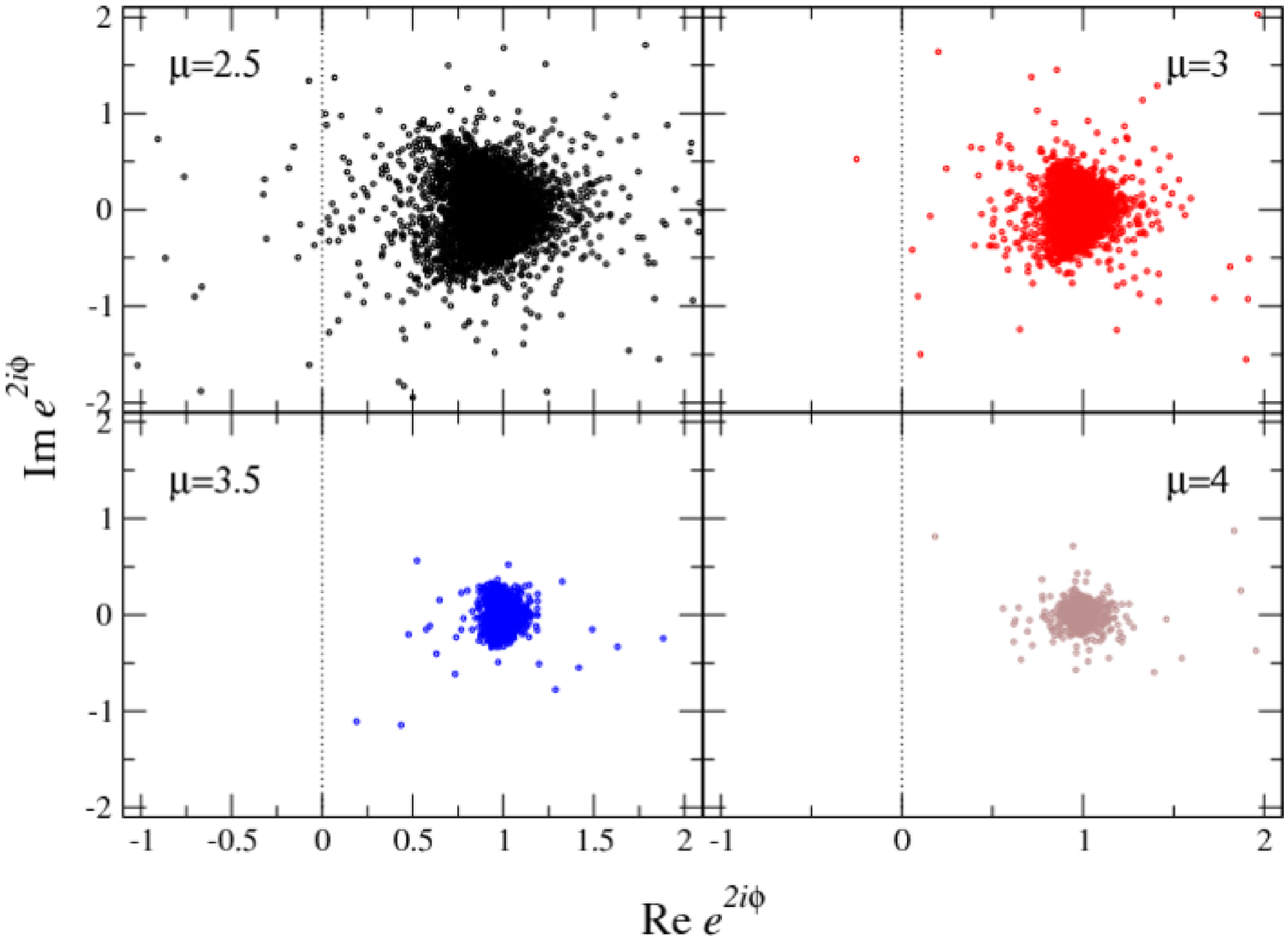,width=9.5cm}
\end{center}
 \caption{Scatter plot of $e^{2i\phi} = \det M(\mu)/\det M(-\mu)$ during 
the Langevin evolution, for $\beta=1$, $\kappa=1/2$ and $0.25\leq \mu\leq 
4$.} 
\label{fig:scatsu3}
\end{figure}

As in the U(1) model, we study the phase of the determinant in the form
\be
 \bra e^{2i\phi} \ket = \left\bra \frac{\det M(\mu)}{\det M(-\mu)} 
\right\ket.
\ee
 At zero chemical potential, the ratio is 1. Due to the SU(3) structure, 
however, the behaviour at large $\mu$ is qualitatively different. We 
find
 \bea
\lim_{\mu\to \infty} \det M(\mu) =&&\hm  \kappa^3e^{3\mu}
 \left[ 1 + 3 e^{-\mu}\left(\kappa+\kappa^{-1}\right) P^{-1} 
+ {\cal O}(e^{-2\mu}) \right],
\\
\lim_{\mu\to \infty} \det M(-\mu) =&&\hm \kappa^3e^{3\mu}
 \left[ 1 + 3 e^{-\mu}\left(\kappa+\kappa^{-1}\right) P 
+ {\cal O}(e^{-2\mu}) \right],
\eea
such that
\be
\lim_{\mu\to \infty} \frac{\det M(\mu)}{\det M(-\mu)} = 1 + 3 e^{-\mu}
\left(\kappa+\kappa^{-1}\right) \left( P^{-1} - P\right) + {\cal 
O}(e^{-2\mu}).
\ee
 As a result the average phase goes to 1 at large $\mu$ and not towards 0 
as in the U(1) model.\footnote{This difference can be traced back to Eq.\ 
(\ref{eq:dettr}).} 
Therefore we expect the sign problem to become exponentially small in 
the saturation regime at large $\mu$.

We have computed the average phase factor and the results are shown in 
Fig.\ \ref{fig:detdetstarsu3} (left). The lines are again the `exact' 
results. As is clear from this plot, the sign problem is quite mild for all 
values of $\mu$, since the maximal deviation from 1 is less than 15\%. In 
Fig.\ \ref{fig:detdetstarsu3} (right) the ratio $\bra\det M(-\mu)/\det 
M(\mu)\ket=Z(-\mu)/Z(\mu)$ is shown. Here we observe a small but systematic 
deviation from 1, more pronounced at smaller $\beta$ and intermediate 
$\mu$. However, we found that the deviation from 1 is reduced when 
continuing the Langevin evolution to larger and larger times. As in the 
U(1) model, this observable is the most sensitive and slowest converging 
quantity.

Scatter plots of the phase are presented in Fig.\ \ref{fig:scatsu3}. At 
small chemical potential (top figure), there appears to be a similar 
structure as in the U(1) model, although not as pronounced. In the 
intermediate region (middle), the distribution is wider. At large $\mu$ 
(bottom), the distribution becomes narrow again, centered around $(1,0)$.

\section{QCD at finite chemical potential}
\label{sec:polyakov}
\setcounter{equation}{0}

\subsection{Hopping expansion}

In this section we leave the one link models behind and consider QCD at 
chemical potential. The SU(3) gauge field contribution to the euclidean 
lattice action is\footnote{We write $U^{-1}$ rather than $U^\dagger$; see 
footnote \ref{fn1}.}
\be
S_B[U] = -\beta \sum_x 
\mathop{\sum_{\mu,\nu}}_{\mu<\nu} 
\left(\frac{1}{6}\left[ \Tr 
U_{x,\mu\nu}+ \Tr U_{x,\mu\nu}^{-1}\right]-1\right),
\ee
with $\beta = 6/g^2$.
The plaquettes are defined as
\be
 U_{x,\mu\nu} = U_{x,\mu} U_{x+\hat\mu,\nu} U^{-1}_{x+\hat\nu,\mu} 
 U^{-1}_{x,\nu},
\ee
and
\be 
 U_{x,\mu\nu}^{-1} = U_{x,\nu\mu}.  
\ee
 The fermion matrix $M$ for Wilson fermions was already given in Eq.\ 
(\ref{eqMQCD}). The $\gamma$ matrices satisfy
$\gamma_\mu^\dagger = \gamma_\mu$ and $\gamma_\mu^2=\id$.
We use periodic boundary conditions in space and antiperiodic boundary 
conditions in the euclidean time direction; the temperature and 
the number of time slices $N_\tau$ are related as $T=1/N_\tau$ (the 
lattice spacing $a\equiv 1$). The fermion matrix obeys
\be
 M^\dagger(\mu) = \gamma_5 M(-\mu) \gamma_5,
\ee 
such that the determinant obeys Eq.\ (\ref{eqdetsym}).

We consider the heavy quark limit, where all spatial hopping terms are 
ignored and only the temporal links in the fermion determinant are 
preserved. We write therefore
\bea
\nn
\det M \approx &&\hm \det \left[ 1
-\kappa \left( e^\mu \Gamma_{+4} U_{x,4}T_{4} +
e^{-\mu}\Gamma_{-4} U_{x,4}^{-1} T_{-4} \right) \right]\\
\nn
= &&\hm \det \left( 1 - 2\kappa e^\mu U_{x,4} T_{4} \right)^2
\det\left( 1 - 2\kappa e^{-\mu} U_{x,4}^{-1} T_{-4} \right)^2\\
\label{eq:detMh}
= &&\hm \prod_{\xv} 
\det\left( 1 + h e^{\mu/T} \cP_{\xv} \right)^2
\det\left( 1 + h e^{-\mu/T} \cP_{\xv}^{-1} \right)^2,
\eea
where we defined $h= (2\kappa)^{N_\tau}$ and the (conjugate) Polyakov 
loops are
\be
 \cP_\xv = \prod_{\tau=0}^{N_\tau-1} U_{(\tau,\xv),4},
\;\;\;\;\;\;\;\;\;\;\;\;\;\;\;\;
 \cP_\xv^{-1} = \prod_{\tau=N_\tau-1}^{0} U_{(\tau,\xv),4}^{-1}.
\ee
 In the final line of Eq.\ (\ref{eq:detMh}) the determinant refers to 
colour space only. The $+$ sign appears because of the antiperiodic 
boundary conditions.

This approximation is motivated by the Heavy Dense Model considered e.g.\ 
in Refs.\ 
\cite{Bender:1991gn,Engels:1999tz,Blum:1995cb,Aarts:2001dz,Hofmann:2003vv,Fukushima:2006uv,DePietri:2007ak}, 
in which the limit
\be
 \kappa\to 0, 
\;\;\;\; \;\;\;\; 
\mu\to \infty, 
\;\;\;\; \;\;\;\; 
\kappa e^\mu \;\;
\mbox{fixed},
\ee
 was taken. However, here also the backward propagating quark, with the 
inverse Polyakov loop, is kept in order to preserve the relation 
(\ref{eqdetsym}).

Using Eq.\ (\ref{eq:dettr}), the determinant can now be written as
\be
\det M = e^{-S_F},
\;\;\;\;\;\;\;\;\;\;\;\;
S_F = -\sum_\xv \left(  2\ln {\cal M}^{(q)}_\xv  +2\ln {\cal M}^{(\bar 
q)}_\xv \right) ,
\ee
with the quark and anti-quark contributions
\bea
{\cal M}^{(q)}_\xv = &&\hm
1+ 3h e^{\mu/T} P_{\xv} +  3h^2 e^{2\mu/T} P_{\xv}^{-1} + h^3 e^{3\mu/T},
\\
{\cal M}^{(\bar q)}_\xv = &&\hm
1+ 3h e^{-\mu/T} P_{\xv}^{-1} +  3h^2 e^{-2\mu/T} P_{\xv} 
+ h^3 e^{-3\mu/T},
\eea
where
\be
 P_\xv = \frac{1}{3}\Tr\cP_\xv,
\;\;\;\;\;\;\;\;\;\;\;\;
 P_\xv^{-1} = \frac{1}{3}\Tr\cP_\xv^{-1}.
\ee
The density is given by
\be 
\bra N\ket = \sum_\xv \bra n_\xv \ket = T\frac{\partial\ln Z}{\partial\mu},
\ee
and we find
\bea
 n_\xv = &&\hm
2T \frac{\partial \ln {\cal M}^{(q)}_\xv}{\partial\mu} + 
2T \frac{\partial \ln {\cal M}^{(\bar q)}_\xv}{\partial\mu}
\nn \\
 = &&\hm 
6\frac{h e^{\mu/T} P_\xv + 2h^2 e^{2\mu/T} P^{-1}_\xv + h^3e^{3\mu/T} }
{{\cal M}^{(q)}_\xv}
\nn \\
&&\hm - 
6\frac{h e^{-\mu/T} P^{-1}_\xv + 2h^2 e^{-2\mu/T}P_\xv + h^3e^{-3\mu/T} 
}{{\cal M}^{(\bar q)}_\xv}.
\eea
At zero chemical potential, the density vanishes while at large 
$\mu$ the density $n_\xv \to 6$, the maximal numbers of quarks on a
site.

\subsection{Complex Langevin dynamics}

The implementation of the Langevin dynamics follows closely the one 
discussed in the previous section on the SU(3) one link model.
We denote $U_{x,\mu}(\theta+\eps) = U'_{x,\mu}$ 
and $U_{x,\mu}(\theta) = U_{x,\mu}$, and consider the process
\be
 \label{eq:flang}
 U'_{x,\mu} = R_{x,\mu}\, U_{x,\mu},
\;\;\;\;\;\;\;\;\;\;\;\;
 R_{x,\mu} = \exp\left[i\lambda_a\left(\eps K_{x\mu a} +\sqrt{\eps} 
\eta_{x\mu a} \right)\right], 
\ee
with the noise satisfying
\be
 \bra \eta_{x\mu a}\ket = 0, \;\;\;\; \;\;\;\; \;\;\;\;
 \bra \eta_{x\mu a}\eta_{y\nu b} \ket = 2 
\delta_{\mu\nu}\delta_{ab}\delta_{xy}.
\ee
The drift term is 
\be
 K_{x\mu a}  = -D_{x\mu a} S[U].
\ee
Differentiation is defined as
\be
 D_{x\mu a} f(U) = \frac{\partial}{\partial\alpha} f\left( e^{i\alpha 
\lambda_a} U_{x,\mu}\right)\Big|_{\alpha=0}.
\ee
The drift term is written as
\be
 K_{x\mu a} =  K_{x\mu a}^B + K_{x\mu a}^F,
\ee 
with the bosonic contribution
\bea
K_{x\mu a}^B = &&\hm -D_{x\mu a} S_B[U] 
\nn\\
= &&\hm i\frac{\beta}{6} \sum_{\nu\neq\mu}\Tr\!\left[
\lambda_a U_{x\mu} C_{x,\mu\nu} 
- \overline{C}_{x,\mu\nu} U_{x\mu}^{-1}\lambda_a \right],
\eea
where
\bea
C_{x,\mu\nu} =
U_{x+\hat\mu,\nu} U^{-1}_{x+\hat\nu,\mu}  U^{-1}_{x,\nu} + 
U^{-1}_{x+\hat\mu-\hat\nu,\nu}  U^{-1}_{x-\hat\nu,\mu} U_{x-\hat\nu,\nu},
\\
\overline{C}_{x,\mu\nu} =
U_{x,\nu} U_{x+\hat\nu,\mu}  U^{-1}_{x+\hat\mu,\nu} + 
U^{-1}_{x-\hat\nu,\nu}  U_{x-\hat\nu,\mu} U_{x+\hat\mu-\hat\nu,\nu}.
\eea
The fermionic contribution is
\be
K_{x\mu a}^F = -D_{x\mu a} S_F[U] = \delta_{\mu 4} K^F_{xa},
\ee
with
\be
K_{xa}^F = 
6\frac{h e^{\mu/T} D_{xa}P_{\xv} + h^2 e^{2\mu/T} D_{xa}P_{\xv}^{-1} }{
{\cal M}^{(q)}_{\xv} } 
+6\frac{h e^{-\mu/T} D_{xa}P_{\xv}^{-1} + h^2 e^{-2\mu/T} D_{xa}P_{\xv}}{
{\cal M}^{(\bar q)}_{\xv} }.
\ee
 The derivatives are
\bea
 D_{xa}P_{\xv} = &&\hm D_{(\tau\xv)a}P_{\xv} = \frac{i}{3} \Tr
\prod_{\tau'=0}^{\tau-1}U_{(\tau'\xv)4} 
\lambda_a \prod_{\tau''=\tau}^{N_\tau-1}U_{(\tau''\xv)4}, 
\\
 D_{xa}P_{\xv}^{-1} = &&\hm D_{(\tau\xv)a}P_{\xv}^{-1} = -\frac{i}{3} \Tr
\prod_{\tau'=N_\tau-1}^{\tau}U_{(\tau'\xv)4}^{-1} 
\lambda_a \prod_{\tau''=\tau-1}^{0}U_{(\tau''\xv)4}^{-1}. 
\eea
 We have solved Eq.\ (\ref{eq:flang}) numerically. A detailed analysis is 
postponed to a future publication; here we present some results for 
illustration purposes.
We have used the temporal gauge, where only the last 
link differs from the identity,
\be
 U_{(N_\tau-1,\xv)4} = V_{\xv}, 
\;\;\;\;\;\;\;\;\;\;\;\;
 U_{(\tau,\xv)4} = \id \;\;\;\;(\tau\neq N_\tau-1).
\ee
 To simplify the exponentiation we use the following updating factor in 
Eq.\ (\ref{eq:flang}),
\be
{\tilde R}_{x,\mu} = \prod_{a={\rm Perm}(1,..,8)}
e^{i\lambda_a\left(\eps K_{x\mu a} +\sqrt{\eps}\eta_{x\mu a} \right) },
\ee
 with random ordering from sweep to sweep and where $K_{x\mu a}$ is 
complex. $R$ and ${\tilde R}$ only differ by terms of order $\epsilon^2$, 
which is the general systematic error of the Langevin algorithm. For the 
results shown here, we have employed Langevin stepsize $\eps=2\times 
10^{-5}$ and 50000 iterations of 50 sweeps each, using ergodicity to 
calculate averages.
 Runaway trajectories have practically been eliminated by monitoring the 
drift and using adaptive step size.
 The lattice has size $4^4$, with $\beta=5.6$, $\kappa=0.12$. We have 
studied chemical potentials in the range $0.5\leq \mu\leq 0.9$, using 
$N_f=3$ fermion flavours.

\begin{figure}[t]
\begin{center}
\epsfig{figure=plot_fsu3_rePPcc_Nt4_ch2.eps,height=5cm}
\epsfig{figure=plot_fsu3_redensity_Nt4_ch2,height=5cm}
\end{center}
 \caption{ Real part of the Polyakov loop $\bra P\ket$ and the conjugate 
Polyakov loop $\bra P^{-1}\ket$ (left) and the density $\bra n\ket$ 
(right) as a function of $\mu$ at $\beta=5.6$, $\kappa=0.12$ on 
a $4^4$ lattice, with $N_f=3$ flavours.
}
 \label{fig:polyfsu3}
\begin{center}
\epsfig{figure=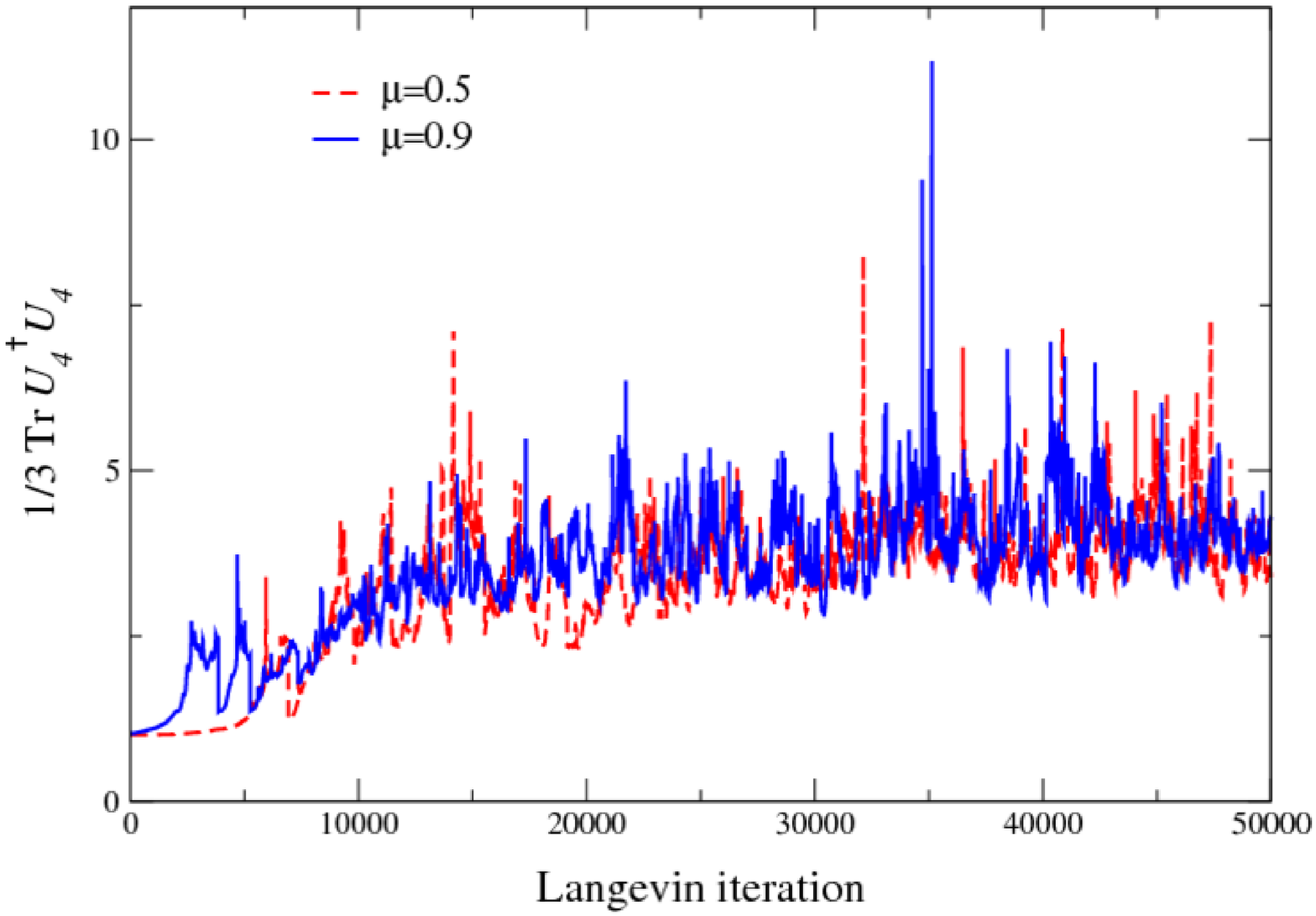,height=5cm}
\end{center}
 \caption{Deviation from SU(3): $\Tr U_4^\dagger U_4/3$ during the Langevin 
evolution, for $\mu=0.5$ and $0.9$.
}
 \label{fig:truudaggerfsu3}
\end{figure}

In Fig.\ \ref{fig:polyfsu3} we present the real part of the Polyakov loop 
and the conjugate Polyakov loop (left) and the density (right). These 
results appear consistent with those obtained in Ref.\ 
\cite{DePietri:2007ak} using reweighting techniques, although at this 
level of the study both statistics and thermalization are not yet 
optimal. Nevertheless we clearly see that at $\mu=0.5$ the system is in 
the low-density ``confining" phase whereas for larger $\mu$ the density 
increases rapidly and both the direct and the conjugate Polyakov loops 
become nonzero, indicating ``deconfinement".

 The deviation from SU(3) during the complex Langevin evolution is shown in 
Fig.\ \ref{fig:truudaggerfsu3}, using $\Tr  U_4^\dagger U_4/3$ as the 
observable. After the initial thermalization stage, this quantity 
fluctuates around $3.5>1$. The fluctuations are similar for all values of 
the chemical potential we considered. Using spatial links $U_{i}$ 
rather than $U_4$ gives a similar result.

\begin{figure}[t]
\begin{center}
\epsfig{figure=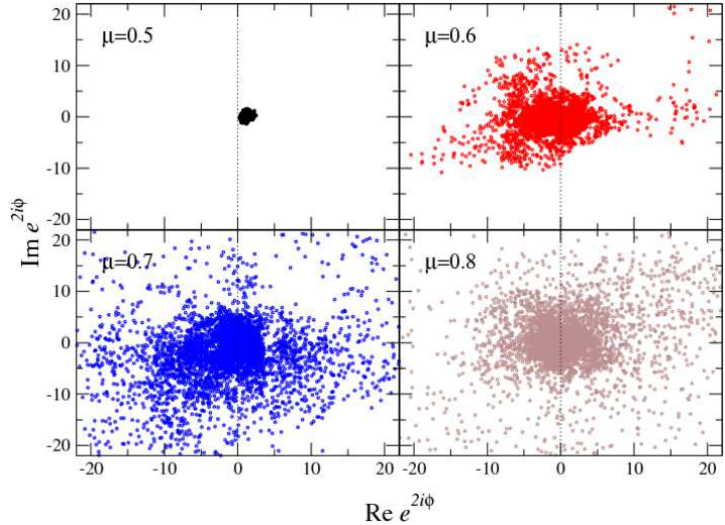,width=9.5cm}
\end{center}
 \caption{Scatter plot of $e^{2i\phi} = \det M(\mu)/\det
M(-\mu)$ during the Langevin evolution for various values of $\mu$ at 
$\beta=5.6$, $\kappa=0.12$ on a $4^4$ lattice. 
}
\label{fig:scatfsu3}
\end{figure}

\subsection{Phase of the determinant}

We study the phase of the determinant as before. Scatter plots during the 
Langevin evolution are shown in Fig.\ \ref{fig:scatfsu3}, for $\mu=0.5, 
0.6, 0.7$, and $0.8$. At the smallest value of $\mu$, the average phase 
factor is close to one; for the real part we find $0.91\pm 0.28$, while 
the imaginary part is consistent with zero ($0.009\pm 0.39$). At the 
larger values shown here, the distribution immediately becomes very wide 
and the average phase factor is consistent with zero (but with a large 
error). Note that the scale is very different compared to the one link 
model. Such an (apparently) abrupt change in the average phase factor when 
moving from a low-density to a high-density phase is somewhat reminiscent 
of what is found in random matrix studies, see e.g.\ Refs.\ 
\cite{Splittorff:2006fu,Han:2008xj,Splittorff:2005wc}.

At large chemical potential the average phase factor approaches 1 again. 
This follows from the behaviour of the determinant, 
\bea
\lim_{\mu\to \infty} \det M(\mu) =&&\hm \prod_\xv h^3e^{3\mu/T}
 \left[ 1 + 3 e^{-\mu/T}\left(h+h^{-1}\right) P_\xv^{-1} 
+ {\cal O}(e^{-2\mu/T}) \right], \;\;\;\;
\\
\lim_{\mu\to \infty} \det M(-\mu) =&&\hm \prod_\xv h^3e^{3\mu/T}
 \left[ 1 + 3 e^{-\mu/T}\left(h+h^{-1}\right) P_\xv 
+ {\cal O}(e^{-2\mu/T}) \right], \;\;\;\;
\eea
such that 
\be
\lim_{\mu\to \infty} \frac{\det M(\mu)}{\det M(-\mu)} = 1 + 3 e^{-\mu/T}
\left(h+h^{-1}\right) \prod_\xv \left( P_\xv^{-1} - P_\xv \right) + {\cal 
O}(e^{-2\mu/T}).
\ee
However, the values of the chemical potential we consider are not in 
that saturation region.

\section{Summary and outlook}
\label{sec:outlook}
\setcounter{equation}{0}

We have considered stochastic quantization for theories with a complex 
action due to finite chemical potential, and applied complex Langevin 
dynamics to U(1) and SU(3) one link models and QCD in the hopping 
expansion. In the latter, the full gauge dynamics is preserved but the 
fermion determinant is approximated. In all cases the complex determinant 
satisfies $\det M(\mu)=[\det M(-\mu)]^*$, as is the case in QCD. We 
studied the (conjugate) Polyakov loops, the density and the phase of the 
determinant. In the one link models excellent agreement between the 
numerical and exact results was obtained, for all values of $\mu$ ranging 
from zero to saturation.

In the one link models the phase of the determinant was studied in 
detail.  Even when the phase factor varies significantly during the 
Langevin evolution, its distribution is well-defined and expectation 
values can be evaluated without problems.  The sign problem does not 
appear to be an obstruction here. In QCD in the hopping expansion, first 
results indicate that the behaviour of the average phase factor changes 
abruptly when moving from the low-density to a high-density phase. 
Nevertheless, other observables (Polyakov loop, density) are still under 
control, even at larger chemical potential.

In the case of the U(1) model, we found strong hints why the sign problem 
does not appear to affect this method. We found that important features 
of classical flow and classical fixed points are largely independent of 
the chemical potential. The presence of $\mu$ changes the complex 
Langevin dynamics only quantitatively but not qualitatively, even when 
the average phase factor of the determinant becomes very small. Moreover, 
a study of the complex Fokker-Planck equation shows that all nonzero 
eigenvalues are real and positive, also in the presence of a nonzero 
chemical potential. An open question concerns the relationship between 
the stationary solution of the complex Fokker-Planck operator and its 
real counterpart. The structure in the U(1) model responsible for these 
results is related to symmetry properties of the action and the 
determinant. Therefore, it may be envisaged that they carry over to the 
more complicated cases.

There are many directions into which this work can be extended. In the 
U(1) model, considerable insight could be obtained (semi-)analytically. It 
will be interesting to extend this analysis to more complicated theories. 
It will also be useful to perform further tests of the method in other 
simple models sharing relevant features with QCD at finite $\mu$. 
Concerning QCD in the hopping expansion, for which we presented first 
results here, a more systematic study stays ahead. One way to test the 
approach is to also perform (real) Langevin dynamics at imaginary chemical 
potential, which goes smoothly and without runaway and convergence 
problems, and perform an analytical continuation. Finally, it will be 
interesting to apply this method to QCD both at large density as well as 
in the region of small chemical potential around the crossover 
temperature. Here it may shed light on the (non)existence of the critical 
point, in a setup which is manifestly independent from the other 
approaches available in this region. It should be noted that the 
Langevin method only requires the derivative of the determinant, and not 
the determinant itself.


 \vspace*{0.5cm}
 \noindent
 {\bf Acknowledgments.}
 We are indebted to Erhard Seiler who collaborated during part of 
this work and contributed many important insights.
G.A.\ thanks Simon Hands, Biagio Lucini and Asad Naqvi for 
discussion.
 This work was initiated while both authors were participating in the 
programme on {\em Nonequilibrium Dynamics in Particle Physics and 
Cosmology} at the Kavli Institute for Theoretical Physics in Santa 
Barbara and supported in part by the National Science Foundation 
under Grant No.\ PHY05-51164.
We thank the organizers for this opportunity. G.A.\ wants to 
thank the organizers of the programme on {\em New Frontiers in QCD 2008: 
Fundamental Problems in Hot and/or Dense Matter} at the Yukawa Institute 
for Theoretical Physics in Kyoto where part of this work was carried out.
I.-O.\ S.\ thanks the Max-Plank-Institute (Werner Heisenberg Institute)
Munich for repeated hospitality during which work for this study was
carried out.
 G.A.\ is supported by an STFC Advanced Fellowship.

\appendix
\renewcommand{\theequation}{\Alph{section}.\arabic{equation}}

\section{Fokker-Planck equation in Minkowski time}
\setcounter{equation}{0}

In Refs.\ \cite{Berges:2005yt,Berges:2006xc,Berges:2007nr} stochastic
quantization was applied to nonequilibrium quantum dynamics in real
time.  For completeness, we give here the analysis of the corresponding
complex Fokker-Planck equation for the one link U(1)
model.\footnote{This Appendix is partly based on Ref.\ \cite{Seiler}.}

Consider the following partition function in Minkowski time, 
\be
 Z_p = \int_{-\pi}^{\pi}\frac{dx}{2\pi}\, e^{iS_p},
\;\;\;\;\;\;\;\;\;\;\;\;\;\;\;\;
 S_p = \beta \cos x +px.
\ee
 The term $px$, with $p$ integer, is a reweighting term, used to 
stabilize the Langevin dynamics (see also Ref.\ \cite{Berges:2007nr}). 
The Fokker-Planck equation for the (complex) distribution $P_p$ reads
 \bea
 \frac{\partial}{\partial \theta} P_p(x,\theta) = &&\hm 
\frac{\partial}{\partial x} \left(  \nu\frac{\partial}{\partial x}
 -i\frac{\partial S_p}{\partial x} \right) P_p(x,\theta) 
\nn \\
= &&\hm \nu P_p''(x,\theta) +
 i\left[ \beta\sin (x) -p\right] P_p'(x,\theta) + 
 i\beta\cos (x) P_p(x,\theta).
\eea
 Here $\nu$ is essentially the normalization of the noise: for $\nu=1$ we 
have the full quantum case, for $\nu=0$ we obtain the classical evolution 
(which is equivalent to taking $\beta$ and $p$ to $\infty$).

To continue we discretize $x$ as $x_l=2\pi l/N$, with $-(N-1)/2 \leq l\leq 
N/2$, and define the modes as
\bea
{\tilde P}_p (n,\theta) =&&\hm \frac{1}{N}\sum_{l=-(N-1)/2}^{N/2} 
e^{inx_l} P_p(x_l,\theta), \\
P_p(x_l,\theta) =&&\hm \sum_{n=-(N-1)/2}^{N/2}
e^{-inx_l} {\tilde P}_p (n,\theta).
\eea 
The Fokker-Planck equation for the modes $\tilde P_p(n,\theta)$ then reads
\bea
\frac{\partial}{\partial \theta} {\tilde P}_p(n,\theta) = 
&&\hm
- \left[\nu\frac{N^2}{\pi^2}\sin^2(k_n/2) + 
 p\frac{N}{2\pi}\sin (k_n)\right]{\tilde P}_p(n,\theta) \nn \\
&&\hm + i\frac{\beta}{2}\frac{N}{2\pi}
\left[ \sin(k_{n+1}) {\tilde P}_p(n+1,\theta) - 
\sin(k_{n-1}) {\tilde P}_p(n-1,\theta)\right] \nn \\
&&\hm -i\frac{\beta}{2}
\left[ {\tilde P}_p(n+1,\theta) - {\tilde P}_p(n-1,\theta)\right],
\eea
where $k_n=2\pi n/N$.
For small $n/N$ this reduces to
\be
\frac{\partial}{\partial \theta} 
{\tilde P}_p(n,\theta) = -(\nu n^2+pn) {\tilde P}_p(n,\theta) 
+ \frac{n}{2}i\beta \left[ {\tilde P}_p(n+1,\theta) - {\tilde 
P}_p(n-1,\theta) \right],
\label{e.fpef}
\ee
 which can be obtained directly from the continuum Fokker-Planck equation 
before discretizing $x$.  Extension to general (complex) $\beta=\beta_R + 
i\beta_I$ and $p=p_R+ip_I$ is straightforward. In the following no explicit 
$\nu$ means $\nu=1$.

From averages with the distribution $P$ we obtain
\bea
 \langle e^{iqx}\rangle_p =&&\hm 
 \frac{\int_{-\pi}^{\pi} dx\,e^{iqx}P_p(x,\theta)}{\int_{-\pi}^{\pi} 
 dx\,P_p(x,\theta)} 
 = 
 \frac{{\tilde P}_p (q,\theta)}{{\tilde P}_p (0,\theta)},
 \label{e.fpap}\\
\langle e^{iqx}\rangle_0 =&&\hm 
 \frac{\int_{-\pi}^{\pi} dx\,e^{iqx}P_0(x,\theta)}{\int_{-\pi}^{\pi} 
 dx\,P_0(x,\theta)} 
 = 
 \frac{\int_{-\pi}^{\pi} dx\,e^{i(q-p)x}P_p(x,\theta)}{\int_{-\pi}^{\pi} 
 dx\,e^{-ipx}P_p(x,\theta)} 
 = 
 \frac{{\tilde P}_p (q-p,\theta)}{{\tilde P}_p (-p,\theta)}.
  \label{e.fpap0}
\eea
 Notice that $p$ and $q$ (both integers) are interchangeable. This implies 
that simulations can be performed at $p=0$, while Eqs.\ (\ref{e.fpap}) -- 
(\ref{e.fpap0}) can be used to obtain $P_p$ for any $p$.

\begin{figure}[t]
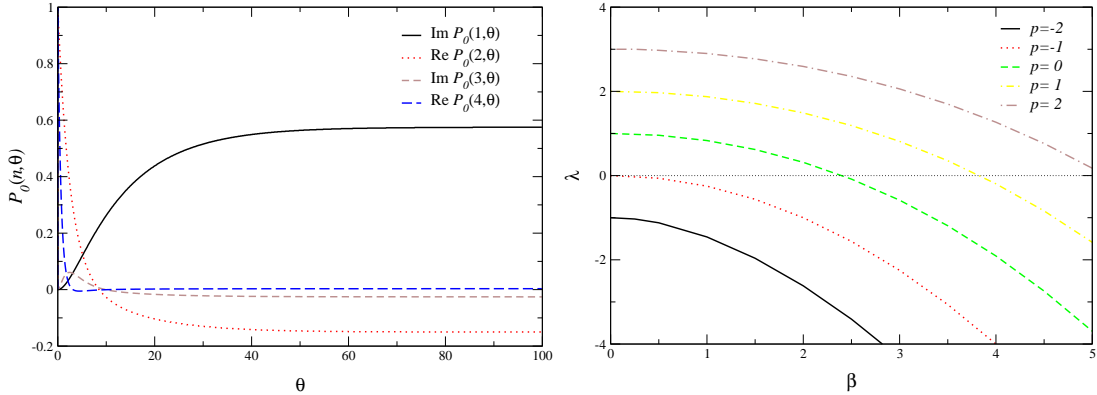

\begin{center}
\epsfig{figure=fp_b1_mink.eps,height=5.2cm}
\epsfig{figure=fokker_cmplxbeta_p.eps,height=5.2cm}
\end{center}
 \caption{Complex Fokker-Planck equation for Minkowski dynamics.
Left: Langevin time dependence of the modes $\tilde P_0(n,\theta)$ for 
various values of $n$, with $\beta=1$, $p=0$.
Right: Smallest nonzero eigenvalue of the complex Fokker-Planck 
equation as a function of $\beta$ for various values of the 
reweighting parameter $p$.
 }
 \label{fig:fpm}
\end{figure}

 Similar to the procedure of Sec.\ \ref{sec:fpeq}, we have solved the 
complex Fokker-Planck equation numerically for the modes $\tilde 
P_p(n,\theta)$. In contrast to the finite $\mu$ case, these modes are now 
complex in general.  For example, for even (odd) $p$ even (odd) modes are 
real and odd (even) modes are imaginary, in agreement with the symmetries 
of the action.  We further find that for $p>0$ the solution for positive 
modes converges correctly, but negative modes diverge, and vice versa. The 
numerical solution for $p=0$ converges quickly to the values determined by 
$e^{iS_{0}}$, when $0\leq \beta \lesssim 2.3$.\footnote{Notice that the 
partition function $Z_0$ has a zero at $\beta$ near $2.4$.} In Fig.\ 
\ref{fig:fpm} (left) we show the Langevin time dependence of some modes 
when $\beta=1$, $p=0$.  Using the asymptotic values for the modes and 
Eqs.\ (\ref{e.fpap}) -- (\ref{e.fpap0}), we can reconstruct $P_p(x)$, 
which agrees nicely with $e^{iS_p(x)}$.  The Langevin simulation itself 
also yields good results for the expectation values when $p\neq 0$ or at 
$p=0$, provided reweighting from $p \neq 0$ is used. For a more thorough 
discussion of the conditions for convergence of the Langevin simulation, 
see Ref.\ \cite{Berges:2007nr}.

Again further insight can be obtained from the eigenvalues, determined by 
the eigenvalue equation
\be
 \label{eq:FPm}
  n(n+p) \tilde P_p(n) + \frac{n}{2} i\beta \left[ \tilde P_p(n+1) - \tilde 
P_p(n-1) \right]
= \lambda \tilde P_p(n).
\ee
 If $\lambda\neq 0$, $P_p(0)$ vanishes and the sequences for $n>0$ and
$n<0$ split again in two. Positive and negative $n$ are related by
changing the sign of $p$. In Fig.\ \ref{fig:fpm} we show the smallest
nonzero eigenvalue for the positive $n$ sequence for five values of $p$.
In contrast to the finite $\mu$ case, we find that eigenvalues may be
negative, depending on the value of $p$ and $\beta$, indicating the
possibility of problems with convergence and stability. This
corroborates the above observations.

Finally, one may also study the real Fokker-Planck equation to obtain the 
true probability distribution $\rho_p(x,y,\theta)$ and its modes. For an 
analytic function $O(z)$ we have
\be
 \int \frac{dxdy}{2\pi}\,\rho_p(x,y,\theta)O(x+iy) = \int 
\frac{dx}{2\pi}\,P_p(x,\theta)O(x),
\ee
hence, in particular,
\be
\int \frac{dxdy}{2\pi}\, \rho_p(x,y,\theta) e^{inx - ny} = 
\int dy\, e^{-ny} {\tilde \rho}_p(n,y,\theta) 
=\int \frac{dx}{2\pi}\,P_p(x,\theta) e^{inx} = {\tilde P}_p(n,\theta). 
\label{e.kydis}
\ee
 The expectation values with $\rho_p$ should represent the averages over 
the 
Langevin process itself. Even when the latter converge to the correct 
values, the real Fokker-Planck equation does not show good behaviour, 
however. We thus have the situation that we find agreement between the 
complex Fokker-Planck equation (with the corresponding complex distribution 
$P$) and the actual Langevin process, while the true probability 
distribution $\rho_p$, which formally mediates between the former two, 
appears more difficult to control.



\begin{thebibliography}{10}



\bibitem{Fodor:2001au}
  Z.~Fodor and S.~D.~Katz,
  Phys.\ Lett.\  B {\bf 534} (2002) 87
  [hep-lat/0104001].

\bibitem{Fodor:2001pe}
  Z.~Fodor and S.~D.~Katz,
  JHEP {\bf 0203} (2002) 014
  [hep-lat/0106002].

\bibitem{Fodor:2002km}
  Z.~Fodor, S.~D.~Katz and K.~K.~Szabo,
  Phys.\ Lett.\  B {\bf 568} (2003) 73
  [hep-lat/0208078].

\bibitem{Fodor:2004nz}
  Z.~Fodor and S.~D.~Katz,
  JHEP {\bf 0404} (2004) 050
  [hep-lat/0402006].


\bibitem{Allton:2002zi}
  C.~R.~Allton {\it et al.},
  Phys.\ Rev.\  D {\bf 66} (2002) 074507
  [hep-lat/0204010].

\bibitem{Allton:2003vx}
  C.~R.~Allton, S.~Ejiri, S.~J.~Hands, O.~Kaczmarek, F.~Karsch, 
  E.~Laermann and C.~Schmidt,
  Phys.\ Rev.\  D {\bf 68} (2003) 014507
  [hep-lat/0305007].

\bibitem{Allton:2005gk}
  C.~R.~Allton {\it et al.},
  Phys.\ Rev.\  D {\bf 71} (2005) 054508
  [hep-lat/0501030].

\bibitem{Gavai:2003mf}
  R.~V.~Gavai and S.~Gupta,
  Phys.\ Rev.\  D {\bf 68} (2003) 034506
  [hep-lat/0303013].



\bibitem{de Forcrand:2002ci}
  P.~de Forcrand and O.~Philipsen,
  Nucl.\ Phys.\  B {\bf 642} (2002) 290
  [hep-lat/0205016].

\bibitem{de Forcrand:2003hx}
  P.~de Forcrand and O.~Philipsen,
  Nucl.\ Phys.\  B {\bf 673} (2003) 170
  [hep-lat/0307020].

\bibitem{deForcrand:2006pv}
  P.~de Forcrand and O.~Philipsen,
  JHEP {\bf 0701} (2007) 077
  [hep-lat/0607017].


\bibitem{D'Elia:2002gd}
  M.~D'Elia and M.~P.~Lombardo,
  Phys.\ Rev.\  D {\bf 67} (2003) 014505
  [hep-lat/0209146].




\bibitem{Kratochvila:2005mk}
  S.~Kratochvila and P.~de Forcrand,
  PoS {\bf LAT2005} (2006) 167
  [hep-lat/0509143].

\bibitem{Ejiri:2008xt}
  S.~Ejiri,
  arXiv:0804.3227 [hep-lat].

\bibitem{Fodor:2007vv}
  Z.~Fodor, S.~D.~Katz and C.~Schmidt,
  JHEP {\bf 0703} (2007) 121
  [hep-lat/0701022].


\bibitem{Parisi:1980ys}
  G.~Parisi and Y.~s.~Wu,
  Sci.\ Sin.\  {\bf 24} (1981) 483.

\bibitem{Damgaard:1987rr}
  For a review, see P.~H.~Damgaard and H.~Huffel,
  Phys.\ Rept.\  {\bf 152} (1987) 227.

\bibitem{Parisi:1984cs}
  G.~Parisi,
  Phys.\ Lett.\  B {\bf 131} (1983) 393.
                                                                                                                                        
\bibitem{Klauder:1985a}
   J.~R.~Klauder and W.~P.~Petersen,
  SIAM J.\ Numer.\ Anal.\ 22 (1985) 1153.


\bibitem{Klauder:1985b}
   J.~R.~Klauder and W.~P.~Petersen,
   J.\ Stat.\ Phys.\ 39 (1985) 53.
                                                                                                                                        
                                                                                                                                        
\bibitem{Gausterer:1986gk}
  H.~Gausterer and J.~R.~Klauder,
  Phys.\ Rev.\  D {\bf 33} (1986) 3678.
                                                                                                                                        
                                                                                                                                        



\bibitem{Karsch:1985cb}
  F.~Karsch and H.~W.~Wyld,
  Phys.\ Rev.\ Lett.\  {\bf 55} (1985) 2242.


\bibitem{Ilgenfritz:1986cd}
  E.~M.~Ilgenfritz,
  Phys.\ Lett.\  B {\bf 181} (1986) 327.

\bibitem{Bilic:1987fn}
  N.~Bilic, H.~Gausterer and S.~Sanielevici,
  Phys.\ Rev.\  D {\bf 37} (1988) 3684.

\bibitem{Ambjorn:1986fz}
  J.~Ambjorn, M.~Flensburg and C.~Peterson,
  Nucl.\ Phys.\  B {\bf 275} (1986) 375.


\bibitem{Berges:2005yt}
  J.~Berges and I.~O.~Stamatescu,
  Phys.\ Rev.\ Lett.\  {\bf 95} (2005) 202003
  [hep-lat/0508030].

\bibitem{Berges:2006xc}
  J.~Berges, S.~Borsanyi, D.~Sexty and I.~O.~Stamatescu,
  Phys.\ Rev.\  D {\bf 75} (2007) 045007
  [hep-lat/0609058].

\bibitem{Berges:2007nr}
  J.~Berges and D.~Sexty,
  Nucl.\ Phys.\  B {\bf 799} (2008) 306
  [0708.0779 [hep-lat]].


\bibitem{Hasenfratz:1983ba}
  P.~Hasenfratz and F.~Karsch,
  Phys.\ Lett.\  B {\bf 125} (1983) 308.



\bibitem{Bender:1991gn}
  I.~Bender {\it et al.},
  Nucl.\ Phys.\ Proc.\ Suppl.\  {\bf 26} (1992) 323.

\bibitem{Engels:1999tz}
  J.~Engels, O.~Kaczmarek, F.~Karsch and E.~Laermann,
  Nucl.\ Phys.\  B {\bf 558} (1999) 307
  [hep-lat/9903030].

\bibitem{Blum:1995cb}
  T.~C.~Blum, J.~E.~Hetrick and D.~Toussaint,
  Phys.\ Rev.\ Lett.\  {\bf 76} (1996) 1019
  [hep-lat/9509002].

\bibitem{Aarts:2001dz}
  G.~Aarts, O.~Kaczmarek, F.~Karsch and I.~O.~Stamatescu,
  Nucl.\ Phys.\ Proc.\ Suppl.\  {\bf 106} (2002) 456
  [hep-lat/0110145].

\bibitem{Hofmann:2003vv}
  R.~Hofmann and I.~O.~Stamatescu,
  Nucl.\ Phys.\ Proc.\ Suppl.\  {\bf 129} (2004) 623
  [hep-lat/0309179].

\bibitem{Fukushima:2006uv}
  K.~Fukushima and Y.~Hidaka,
  Phys.\ Rev.\  D {\bf 75} (2007) 036002
  [hep-ph/0610323].

\bibitem{DePietri:2007ak}
  R.~De Pietri, A.~Feo, E.~Seiler and I.~O.~Stamatescu,
  Phys.\ Rev.\  D {\bf 76} (2007) 114501
  [0705.3420 [hep-lat]].


\bibitem{abra}
 Abramowitz, M. and Stegun, I. A. (Eds.), Handbook of Mathematical 
Functions, New York: Dover, p.\ 17, 1972.


\bibitem{Splittorff:2006fu}
  K.~Splittorff and J.~J.~M.~Verbaarschot,
  Phys.\ Rev.\ Lett.\  {\bf 98} (2007) 031601
  [hep-lat/0609076].

\bibitem{Han:2008xj}
  J.~Han and M.~A.~Stephanov,
  arXiv:0805.1939 [hep-lat].

\bibitem{Splittorff:2005wc}
  K.~Splittorff,
  arXiv:hep-lat/0505001.
  
\bibitem{Seiler}
 E.~Seiler and I.-O.~Stamatescu, unpublished (2007).

\end{thebibliography}
\end{document}